\documentclass[conference]{IEEEtran}
\IEEEoverridecommandlockouts
\usepackage{cite}
\usepackage{adjustbox}
\usepackage{graphicx} 
\usepackage{booktabs}
\usepackage{algpseudocode} 
\usepackage{algorithm}
\usepackage{tikz}
\usepackage{amsmath,amssymb,amsfonts}
\usepackage{graphicx}
\usepackage{textcomp}
\usepackage{xcolor}
\usepackage{listings}
\usepackage{multirow}
\usepackage{threeparttable}
\usepackage{capt-of}
\definecolor{lightforestgreen}{rgb}{0.13, 0.55, 0.13} 
\definecolor{lightergray}{rgb}{0.9, 0.9, 0.9} %
\usetikzlibrary{arrows.meta, positioning}
\usepackage[utf8]{inputenc}
\usepackage{tikz}
\usetikzlibrary{shapes.geometric, arrows.meta, positioning, fit, backgrounds, calc}
\usepackage[margin=0.5in]{geometry}
\usepackage{amsmath}

\definecolor{startcolor}{RGB}{144, 238, 144}
\definecolor{processcolor}{RGB}{173, 216, 230}
\definecolor{processcolor2}{RGB}{255, 255, 255}
\definecolor{decisioncolor}{RGB}{255, 255, 224}
\definecolor{retraincolor}{RGB}{252, 184, 139}
\definecolor{retraincolor2}{RGB}{204, 138, 96}
\definecolor{lifecyclecolor}{RGB}{228, 202, 250}
\definecolor{lifecyclecolor2}{RGB}{121, 90, 148}
\definecolor{myblue}{RGB}{173, 216, 230}
\definecolor{myblue2}{RGB}{85, 142, 161}

\tikzstyle{startstop} = [ellipse, minimum width=2cm, minimum height=0.8cm, 
                         text centered, draw=black, fill=startcolor, font=\small\bfseries]
\tikzstyle{process} = [rectangle, rounded corners, minimum width=2.8cm, minimum height=0.9cm, 
                       text centered, draw=black, fill=processcolor2, font=\scriptsize, align=center]
\tikzstyle{decision} = [diamond, minimum width=2.5cm, minimum height=0.8cm, 
                        text centered, draw=black, fill=decisioncolor, font=\scriptsize, aspect=2.5, align=center]
\tikzstyle{window0} = [rectangle, rounded corners, minimum width=3cm, minimum height=1.6cm, 
                       text centered, draw=black, fill=processcolor, font=\tiny, align=center, text width=2.8cm]
\tikzstyle{windowN} = [rectangle, rounded corners, minimum width=3cm, minimum height=1.6cm, 
                       text centered, draw=black, fill=processcolor, font=\tiny, align=center, text width=2.8cm]
\tikzstyle{retrain} = [rectangle, rounded corners, minimum width=2.6cm, minimum height=1.5cm, 
                       text centered, draw=black, fill=retraincolor, font=\tiny, align=center, text width=2.4cm]
\tikzstyle{lifecycle} = [rectangle, rounded corners, minimum width=3cm, minimum height=1.2cm, 
                         text centered, draw=black, fill=lifecyclecolor, font=\tiny, align=center, text width=2.8cm]
\tikzstyle{output} = [rectangle, rounded corners, minimum width=2.2cm, minimum height=0.8cm, 
                      text centered, draw=black, fill=processcolor2, font=\scriptsize, align=center]
\tikzstyle{merge} = [circle, minimum size=0.3cm, fill=black, draw=black]
\tikzstyle{arrow} = [thick,-{Stealth[length=2.5mm]}]
\tikzstyle{arrow_label} = [font=\tiny, midway, fill=white, inner sep=1pt]

\lstdefinestyle{mystyle}{
    backgroundcolor=\color{lightergray},   
    basicstyle=\ttfamily\footnotesize\color{lightforestgreen},  
    frame=none,                      
    breaklines=true,
}
\def\BibTeX{{\rm B\kern-.05em{\sc i\kern-.025em b}\kern-.08em
    T\kern-.1667em\lower.7ex\hbox{E}\kern-.125emX}}
    
\begin{document}

\title{Tracking What Works: Social Engineering Lures as Signatures for Victim-Centric Infostealer Campaign Discovery}

\author{
    \IEEEauthorblockN{Estelle Ruellan}
    \IEEEauthorblockA{\textit{Flare Systems} \\
    Strasbourg, France}
}

\maketitle

\begin{abstract}
Despite advances in antivirus technology, infostealer infections persist at scale, driven primarily by social engineering rather than exploits. Yet, existing campaign detection efforts have focused on technical artifacts, overlooking the social-engineering lures exploiting human behavior and driving successful infections. \\ 
To address this gap, we introduce a dual-framework system that discovers and tracks infostealer campaigns using victim-side artifacts from stealer logs. By treating social engineering lures as campaign signatures, the framework automatically aggregates individual infections into high-fidelity campaigns in near real-time. Recognizing distinct patterns between software and gaming ecosystems, we developed ecosystem-specific pipelines detecting all large and medium campaigns in both ecosystems (one medium gaming campaign was only recovered merged with another), though performance decreased for small campaigns. 
Analysis of 177 real-world campaigns identified by our framework reveals distinct temporal and infrastructure patterns with direct remediation implications. Software campaigns accumulate over 60\% of infections within the first ten days, demanding rapid intervention following detection, while gaming campaigns grow more intense the longer they last, requiring continuous monitoring. \\
Using social engineering lures as campaign signatures provides robustness to infrastructure fragmentation where traditional IoC-based detection fails. 
This work shifts campaign tracking from infrastructure correlation to victim-centric analysis, capturing \textit{why} campaigns succeed rather than only \textit{how} they operate, enabling awareness-driven prevention that addresses the human decision-making dimension absent from existing research.

\end{abstract}


\section{INTRODUCTION}
Infostealers have emerged as one of the most pervasive and impactful forms of malware in today’s threat landscape. By covertly harvesting credentials, session cookies, cryptocurrency wallets, and other sensitive data from infected devices, infostealers enable large-scale account compromise, financial fraud, and ransomware attacks. 

Despite advances in antivirus technologies and malware detection systems, infostealer infections continue to occur at scale, with tens of millions of compromised devices observed globally in recent years. A key reason for this persistence lies in how infostealers spread. Unlike exploit-driven malware, infostealers predominantly rely on social engineering. Users are persuaded to download and run malware disguised as cracked software, gaming tools, productivity applications, or other seemingly legitimate applications, often while being instructed to disable security protections for 'successful' installation.

In this setting, the success of an infostealer campaign is less driven by technical sophistication and more by its ability to exploit human decision-making.

Malware detection research has traditionally focused on identifying malicious activity through static file characteristics, dynamic behavior, network traffic, and infrastructure-level indicators. Parallel efforts in campaign detection have successfully grouped malicious activity by correlating technical artifacts such as domains, IP addresses, URLs, and communication patterns. While these approaches are effective at identifying coordinated activity, they provide limited insight into why campaigns succeed, and very few specifically address infostealers or the social engineering mechanisms that drive their infections.

At the same time, successful infostealer campaigns generate massive amounts of post-infection data. Stealer logs circulate in the millions across underground platforms, implicitly capturing the infection vectors, lures, and deceptive techniques that successfully led to infection (i.e. they worked in real-life scenarios). These artifacts represent a uniquely valuable but largely untapped data source for understanding malware campaigns from the victim’s perspective. Their volume, however, makes manual analysis unrealistic, which has limited their use in both academic and industry research.

Recent advances in Large Language Models (LLMs) have opened new opportunities for processing these artifacts at scale. Prior work has shown that LLMs can extract structured information from infostealer logs, enabling automated analysis of individual infections. What remains missing is a way to move beyond isolated cases and derive campaign-level intelligence from this data.

In this paper, we address this gap by introducing a framework for discovering and tracking infostealer campaigns using victim-side infection artifacts. By leveraging infection themes extracted with an LLM-based pipeline and treating social engineering lures as campaign signatures, our approach enables near real-time campaign identification and analysis. This work shifts the focus of campaign detection toward successful infections, providing new insights into campaign structure, evolution, and the social engineering strategies that drive infostealer success.

\section{PREVIOUS WORK}
    
Malware detection has historically relied on the characterization of malicious intent through either the structural properties of a file or its observed runtime behavior. Existing detection techniques can be broadly categorized into static, signature-based and behavior-based methods that rely on dynamic or memory analysis ~\cite{nataraj_malware_2011, chakravarty_study_nodate, chen_understanding_2019, maniriho_systematic_2024}. 

Static approaches essentially analyze the file without execution, focusing on its structure and content. Signature-based detection remains widely used due to its computational efficiency, as it identifies threats by matching binary patterns against known malware fingerprints. Machine learning classifiers, most notably Random Forest (RF) and Support Vector Machines (SVM), have demonstrated significant success in static contexts~\cite{shokouhinejad2025recent}. However, this reliance on predefined signatures limits its effectiveness when confronting previously unseen or evolving malware variants \cite{nataraj_malware_2011, noauthor_comprehensive_nodate, maniriho_systematic_2024}. 

On the other hand, behavior-based approaches execute the file in a controlled environment, such as a sandbox, and observe its behavior. 
Modern behavioral detection frameworks leverage a diverse range of architectures, including transformer-based models like BERT, Long Short-Term Memory (LSTM) networks, and Convolutional Neural Networks (CNN), alongside graph-based algorithms like Heterogeneous Graph Neural Networks (HetGNN) and probabilistic models such as Hidden Markov Models (HMM), to extract complex temporal and structural patterns from dynamic execution logs~\cite{maganur2025feature, shokouhinejad2025recent}. Although dynamic methods are more resilient to obfuscation, packing, or encryption, they are more computationally demanding and may fail to trigger detection if the malware’s malicious behavior is not activated under the observed conditions \cite{nataraj_malware_2011, chen_understanding_2019}.

With the advent of Large Language Models (LLMs), recent years have seen the emergence of new methods leveraging LLMs for malware detection and analysis ~\cite{zahan_leveraging_2024, rondanini_large_2024, al-karaki_exploring_2024, omar_harnessing_2024, sayed_fine-tuning_2024, sanchez_sanchez_transfer_2024, hossain_malicious_2024}. Research shows the potential of LLMs applied to malware detection in various domains such as classifying malicious code ~\cite{hossain_malicious_2024},  IoT malware ~\cite{omar_harnessing_2024}, analyzing network traffic ~\cite{omar_harnessing_2024}, system call-based malware detection~\cite{sanchez_sanchez_transfer_2024}, and malicious package detection~\cite{zahan_leveraging_2024}.
   
While much of this work focuses on general malware detection, very few studies have explored the use of infection artifacts from victims to detect and analyze malware. In 2025, Ruellan and colleagues used LLMs to process infection screenshots, images captured at the point of compromise and included in infostealer logs, to characterize malware infections. Using gpt-4o-mini, the authors extracted Indicators of Compromise (IoCs), identified infection vectors, and inferred the thematic lure associated with each infection ~\cite{ruellan2025llm}.  

A separate line of research has shifted focus from individual malware detection to the detection and analysis of malware campaigns. In 2015, Zhang and colleagues proposed SMASH, an unsupervised malware campaign detection system that identifies malicious campaigns by mining and correlating groups of related servers using HTTP traces from a large ISP~\cite{zhang2015systematic}. By clustering servers based on shared clients, URI patterns, IP addresses, and Whois information, their approach uncovered 236 malicious campaigns with more than 10,000 malicious servers involved, including zero-day attacks, with low false positives.  

Kwon and colleagues proposed Beewolf in 2016, an unsupervised graph-mining system to detect silent malware delivery campaigns by identifying lockstep behavior between downloaders and payload-hosting domains. Locksteps are synchronized download patterns where groups of downloaders access common domains within short time windows. Using near-biclique patterns in bipartite graphs under temporal constraints on 33.3 million download events, Beewolf detected over 130,000 locksteps comprising 1.4 million campaigns, achieving over 92\% true positive rate with fewer than 5\% false positives~\cite{kwon2016catching}. 

Starov and colleagues (2018) proposed a threat-agnostic approach for detecting and tracking malicious campaigns by exploiting the reuse of third-party web analytics identifiers across websites. Using analytics-ID matching over 18 analytics platforms on large-scale crawls of known malicious URLs, their approach enabled the discovery of more than 11,000 previously unknown malicious domains and the identification of 13 phishing campaigns~\cite{starov2018betrayed}.

Complementary to traffic and infrastructure analysis, researchers have also leveraged the wealth of information contained in threat intelligence reports. In 2018, Zhu and Dumitras introduced Chainsmith, a framework using automated IoC extraction and semantic analysis of threat intelligence reports to analyze and map malware delivery campaign stages~\cite{zhu2018chainsmith}. Chainsmith processed 14,155 online security articles, extracting 24,653 IoCs and identifying 59 campaign groups. It analyzed context words in adjacent sentences to determine campaign stages and used multi-class classifiers to verify IoC roles. Using Chainsmith, the authors established campaign behaviors at each stage of the kill chain, associating them with their respective IoCs~\cite{zhu2018chainsmith}. 

Lastly, Burton and colleagues (2020) identified and tracked malicious spam campaigns by applying graph-based analysis to large-scale email metadata to identify isolated malware campaigns~\cite{burton2020wordythief}. Then, using bipartite graphs, time-series analysis, and DNS correlation they infer shared infrastructure and actor-level behavior. Their approach enabled the discovery and long-term tracking of the WordyThief spam operation, revealing its use of dictionary domain generation algorithms (DDGAs) and a large infrastructure distributing information-stealing malware~\cite{burton2020wordythief}.

This body of research has focused on the detection and grouping of web injection campaigns, malicious communication and attack campaigns~\cite{zhang2015systematic}, silent malware delivery campaigns~\cite{kwon2016catching}, malware delivery campaign stages mined from threat intelligence reports~\cite{zhu2018chainsmith}, email spamming campaigns, including one information-stealing campaign~\cite{burton2020wordythief}, and finally, phishing and malicious campaigns at large~\cite{starov2018betrayed}.

Existing campaign detection research has predominantly focused on technical artifacts, network traffic patterns, infrastructure reuse, URL characteristics, IoCs, domain names, and email metadata, to identify and group malicious activities. 
While these approaches successfully detect campaigns through their technical fingerprints and infrastructure footprints, they fundamentally overlook the initial infection vector that relies on human decision-making: socially engineered lures and baiting techniques that persuade users to voluntarily download and execute malware are mostly absent from the research.

\section{RESEARCH PROBLEM}

The technical focus of existing research assumes that detection and blocking mechanisms (antivirus, blacklists, or network filters) will prevent infections once campaigns are identified. Yet, millions of devices remain infected with infostealers globally, showing that technical approaches, while effective in many cases, are insufficient when human behavior is exploited.

This limitation is particularly critical for infostealer campaigns, where the user is not a passive victim of an exploit but an active participant in their own infection. Infostealer campaigns differentiate themselves less by their payloads or infrastructure than by what they masquerade as, namely the social engineering lures that persuade users to self-infect. Cracked software, gaming tools and productivity applications are not incidental delivery mechanisms but defining characteristics of campaigns themselves. 
Yet, existing threat intelligence standards such as OpenIOC~\footnote{openioc.org} and STIX~\footnote{http://stixproject.github.io/} provide little to no structured support for capturing these lures or the baiting techniques that enable them. As a result, an essential dimension of campaign identity remains largely absent from the literature and operational detection frameworks: how threat actors lure users into downloading the malware and then persuade them into executing it.

\section{PRESENT RESEARCH}

Over the past five years, more than 110 million infostealer logs have been shared across underground platforms. Despite advances in antivirus technology and campaign detection, these infections continue to occur at scale, indicating that infostealers remain highly effective. Preventing such infections requires understanding how they occur in practice: the social engineering tactics, malicious downloads, and deceptive schemes that lead users to self-infect.

Crucially, these successful infections represent the most effective malware in the ecosystem, as they have demonstrably bypassed existing defenses and compromised real devices. Yet, they remain largely unstudied. Post-infection artifacts, such as stealer logs, circulate in the millions across underground channels, forming a rich but untapped data source. Their sheer volume, however, makes manual analysis impractical, and as a result, this reactive perspective on malware has been largely overlooked in both academic and industry research.


The LLM-based pipeline introduced in ~\cite{ruellan2025llm} partially addresses this challenge by automating per-infection analysis, extracting Indicators of Compromise (IoCs) and infection themes at scale. While this significantly accelerates the study of individual infections, the pipeline operates strictly at the infected-device level and does not produce campaign-level intelligence. Campaign identification still requires manual correlation across isolated analyses, which does not scale to the volume of infostealer activity observed in practice. To the best of our knowledge, there exists no framework, methodology, or algorithm capable of identifying and tracking infostealer campaigns at scale using victim-side artifacts, nor one that leverages social engineering lures and baiting techniques as campaign signatures.

Recognizing these limitations, this research introduces a Campaign Discovery and Tracking Framework that builds directly on victim-derived infection artifacts. Leveraging infection themes extracted via the LLM pipeline in ~\cite{ruellan2025llm}, the proposed framework automatically aggregates individual infections into high-fidelity campaigns and tracks their evolution over time in near real-time. By treating social engineering lures as campaign signatures, this work fills a critical gap in the literature on campaign analysis.

Beyond campaign identification, the framework enables the systematic study of campaign characteristics, including longevity, infection intensity, and survival. More broadly, by cataloging and analyzing the lures that have proven successful in real-world infections, this approach opens the door to a prevention paradigm shift: empowering users and defenders through awareness of effective baiting techniques, rather than relying solely on technical controls that can be socially engineered into irrelevance.

The primary objectives of this research are to:
\begin{itemize}
    \item Quantitatively validate the fidelity of identified clusters and campaigns through manual evaluation, demonstrating that the framework detects \textbf{real-world infostealer campaigns}.
    \item Analyze campaign characteristics, including longevity, infection intensity, and survival patterns.
\end{itemize}

\section{METHODOLOGY}
\subsection{Data Collection and Preprocessing}
Infostealer logs were obtained through Flare’s platform. Flare is a dark web monitoring company that maintains a cyber threat intelligence platform by monitoring various online spaces\footnote{https://flare.io/}. 

Our framework processes raw infostealer logs. The data used for this study consisted of 53,896 anonymized logs observed between June 7th and August 12th, 2025, containing detailed information, including victim machine metadata, captured browser history, and a list of files on the system.

First, we focus on extracting the infection theme from the logs. This theme is a concise natural language description of the infection vector (e.g. the name of the file downloaded by the victim). The theme extraction is performed using ~\cite{ruellan2025llm}'s LLM-based pipeline to extract a succinct thematic string and its associated URL, when available.

The output is a dataset where each row has a log id, an infection theme, a URL (when available) and an infection screenshot.

\textbf{Examples of extracted infection themes:}
\begin{itemize}
    \item "GTA V Kiddions Modest Menu undetected 2024"
    \item "Adobe Photoshop CC 2024 crack multilingual"
    \item "ROBLOX Solara executor v3.2 free download"\\
\end{itemize}

\subsection{Embedding Generation and Dimensionality Reduction}

Infection themes were converted into a 1,536-dimensional vector space using OpenAI's \texttt{text-embedding-3-small} model. 

\begin{figure}[h]
    \centering
    \includegraphics[width=\columnwidth]{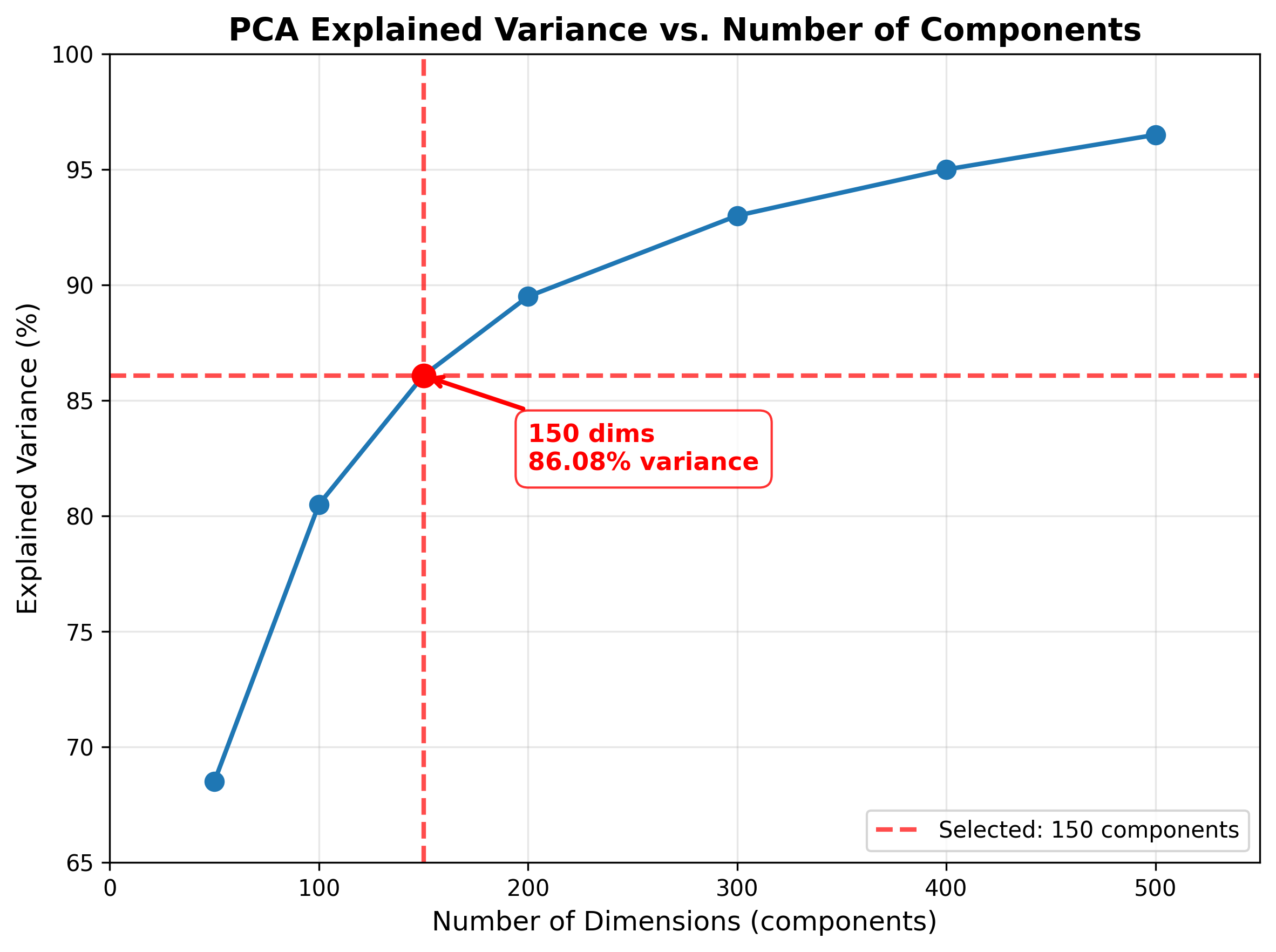}
    \caption{PCA dimensionality reduction analysis. We selected 150 components (red lines) which retain 86.08\% of the original variance while providing significant computational speedup.}
    \label{fig:pca}
\end{figure}

To optimize clustering performance, the embeddings were reduced using Principal Component Analysis (PCA). We selected a final dimension of 150 components, which accounts for $86.08\%$ of the cumulative variance explained (Figure~\ref{fig:pca}).

\subsubsection{Initial Clustering Attempt and Brand Discovery}

An initial attempt to cluster all 53,896 embeddings simultaneously (agnostic to software/game brands) yielded poor results across multiple algorithms (HDBSCAN, K-means, hierarchical clustering) and parameter configurations. Inspection of the resulting clusters revealed that embeddings were grouped primarily by targeted brand rather than by campaign characteristics. For example, all ROBLOX-related infections clustered together regardless of the specific executor or distribution method, while all Adobe-related infections formed a separate cluster.

This observation led to a critical insight: \textbf{effective campaign detection requires brand-specific clustering}. The semantic space of "ROBLOX Valex executor" is fundamentally different from "Adobe Photoshop Multilingual v2.43.6 crack".

For the remaining of this paper, we refer to the targeted product or game (e.g., Adobe, ROBLOX) as a brand, and to the software/gaming split as an ecosystem.

\subsubsection{Brand Classification}

To enable brand-specific analysis, we trained a BERT-based classifier to extract the targeted brand (software or game) from each infection theme. The classifier was fine-tuned on a curated dataset of 27,500 infection themes with manual labels, achieving $\mathbf{93.42\%}$ accuracy (F1=$\mathbf{0.684}$). 

\footnote{Note on Imbalance: The high overall Accuracy ($\mathbf{93.42\%}$) combined with a moderate F1 Score ($\mathbf{0.684}$) indicates a high degree of class imbalance favoring the top-tier brands. Crucially, the model's F1 score demonstrates sufficient accuracy for the top 50 targeted brands, which are the focus of our campaign tracking efforts. This trade-off was accepted as the goal is robust classification for the most active entities, not perfect balance across all brands.}

\textbf{Distribution of Targeted Brands:} Analysis of the 53,896 logs revealed 9,309 unique brands. However, the distribution was highly concentrated, with the top 10 brands responsible for 34.32\% of all infections (Table~\ref{tab:software_distribution}).

\begin{table}[h]
\centering
\caption{Top 10 brands by infection count. These brands account for 34.32\% of all infections in our dataset.}
\label{tab:software_distribution}
\begin{tabular}{lr}
\toprule
\textbf{Brand} & \textbf{Count} \\
\midrule
KMS & 3,049 \\
Grand Theft Auto & 2,623 \\
Microsoft & 2,537 \\
Internet Download Manager & 2,378 \\
ROBLOX & 2,371 \\
Adobe & 2,301 \\
Valorant & 907 \\
Windows & 833 \\
Counter Strike & 800 \\
Fortnite & 697 \\
\midrule
\textbf{Total (Top 10)} & \textbf{18,496} \\
\textbf{Percentage of Dataset} & \textbf{34.32\%} \\
\bottomrule
\end{tabular}
\end{table}

\textbf{CLUSTERING PIPELINE}

A vertical workflow of the Hybrid Rolling HDBSCAN Framework is presented in Figure~\ref{fig:full_workflow_vertical} in the appendix.

\subsection{The Hybrid Rolling HDBSCAN Framework}
\label{sec:hybrid_hdbscan}

Campaign detection is fundamentally a streaming and temporal clustering problem. We considered three clustering algorithms for the campaign clustering. Table~\ref{tab:clustering_comparison} summarizes the strengths and limitations of candidate clustering algorithms for campaign detection.

\begin{table}[h]
\centering
\caption{Pros and cons of candidate clustering algorithms for campaign detection.}
\label{tab:clustering_comparison}
\begin{tabular}{l|p{0.38\columnwidth}|p{0.38\columnwidth}}
\toprule
\textbf{Algorithm} & \textbf{Pros} & \textbf{Cons} \\
\midrule
K-means &
Fast and simple; scalable for static datasets &
Static by design; Requires pre-defined number of clusters; assumes spherical clusters; unsuitable for evolving campaign structures \\

\midrule
DenStream &
Designed for streaming data; supports concept drift &
Computationally prohibitive at scale\\

\midrule
HDBSCAN &
Automatic cluster discovery; density-based; handles arbitrary shapes; scalable &
Static by design; no native support for temporal or incremental tracking \\

\bottomrule
\end{tabular}
\end{table}

K-means is limited to static settings, requires pre-specifying the number of clusters, and assumes spherical cluster geometry, making it unsuitable for identifying unknown and irregularly shaped campaigns. DenStream provides an incremental, streaming-oriented approach; however, preliminary experiments showed that its computational cost is prohibitive at our scale. Given the requirements for scalability (processing thousands of themes daily) and automatic cluster discovery, we selected HDBSCAN as the base algorithm and extended it for temporal tracking. 

Traditional HDBSCAN is static, requiring retraining on the entire dataset, which is computationally expensive and incompatible with tracking campaigns over time. So, we developed a \textbf{Hybrid Rolling HDBSCAN} methodology that adapts static HDBSCAN for incremental, temporal tracking. 

Figure \ref{fig:full_workflow_vertical} in the appendix provides a high-level overview of the complete workflow of the proposed Hybrid Rolling HDBSCAN framework. The algorithm operates on a fixed-size sliding window ($W_{\text{size}}$ days, with a $S_{\text{step}}$ day slide).

\begin{enumerate}
    \item \textbf{Initialization and Incremental Step}:
    \begin{itemize}
         \item \textbf{Window 0 (Initial Training)}: The model is trained on the full $W_{\text{size}}$ dataset using standard HDBSCAN, establishing the initial cluster structure and computing cluster centroids. A \textbf{scale factor} $\sigma$ is computed as the standard deviation of inter-centroid distances, used to normalize all subsequent distance calculations for portability across different embedding spaces.
        \item \textbf{Window $n>0$ (Incremental Assignment)}: New data points are assigned to existing clusters using HDBSCAN's \texttt{approximate\_predict} function, which leverages the trained model's condensed tree structure for efficient assignment without full retraining.
    \end{itemize}
    
     \item \textbf{Noise Reclustering for Novel Campaigns}:
     Points that cannot be assigned to existing clusters become \textbf{noise}. To identify emerging campaigns, we apply \textbf{DBSCAN} exclusively to \textbf{all noise points (old noise + new noise)} within the current window. Any resulting DBSCAN cluster meeting the minimum size threshold is registered as a new cluster with a new global cluster ID. This step is critical to identify new campaigns early on.
     
     \item \textbf{Cluster Lifecycle Management}:\\ 
     To prevent noisy re-alerts and track campaign evolution, clusters transition through five lifecycle states:
    \begin{itemize}
        \item \textbf{NEW:} First detection in current window.
        \item \textbf{ACTIVE:} Currently receiving points in the current window.
        \item \textbf{DORMANT:} No points in current window but retained in memory for up to $M_{\text{dormant}}$ windows. This prevents re-alerting for recurring campaigns.
        \item \textbf{REVIVED:} A dormant cluster that receives points again (detected when a new cluster's centroid matches a dormant cluster's centroid within threshold $\tau_{\text{match}}$).
        \item \textbf{EXPIRED:} Permanently deleted after remaining dormant for more than $M_{\text{dormant}}$ windows.
    \end{itemize}
    
    \item \textbf{Adaptive Retraining Logic}:\\
    The model is retrained only when cluster structure is significantly degraded. We employ a three-tier decision system: \textbf{full retrain} for systemic issues, \textbf{surgical dissolution} for localized contamination, or \textbf{continue incremental} when stable.
    
    \begin{itemize}
    \item \textbf{Centroid Drift Detection:} We compute per-cluster drift rather than average drift to avoid masking localized contamination. For each cluster $i$, the normalized drift is:
    \begin{equation}
        \delta_i = \frac{\|\mathbf{c}_i^{\text{new}} - \mathbf{c}_i^{\text{old}}\|}{\sigma}
    \end{equation}
    where $\sigma$ is the scale factor (standard deviation of inter-centroid distances). A cluster is considered \textbf{drifted} if $\delta_i > \tau_{\text{drift}}$.
    
    \item \textbf{Drift-Based Actions:} Rather than immediately triggering retraining for any drift, we use percentage-based thresholds that adapt to the current cluster count:
    \begin{itemize}
        \item If $< P_{\text{drift}}$ of clusters have drifted: \textbf{Surgical dissolution}, only the drifted clusters are converted to noise and reclustered via DBSCAN, preserving stable clusters.
        \item If $\geq P_{\text{drift}}$ of clusters have drifted: \textbf{Full retrain}, systemic contamination detected.
    \end{itemize}
    
    \item \textbf{Volume-Based Dissolution:} Retraining is also triggered if a \textbf{large cluster} (size $> \tau_{\text{large}}$, default: 50 points) dissolves, or if $\geq P_{\text{dissolve}}$ of clusters dissolve simultaneously due to insufficient points.
    
    \item \textbf{Unassigned Ratio:} If more than $\tau_{\text{unassigned}}$ of new points cannot be assigned to existing clusters, the cluster structure is considered stale and retraining is triggered. 
    
    \item \textbf{Cluster Identity Preservation Post-Retraining:} After full HDBSCAN retraining, new clusters must be matched to old clusters to preserve campaign tracking continuity. We use the Hungarian algorithm to match old and new centroids (including dormant clusters). The matching uses normalized distances:
    \begin{equation}
        d_{\text{normalized}}(c_i, c_j) = \frac{\|c_i - c_j\|}{\sigma}
    \end{equation}
    Matched clusters (normalized distance $< \tau_{\text{match}}$) preserve their global IDs. Dormant clusters that match to new clusters are marked REVIVED. Unmatched new clusters receive new global IDs and are marked NEW.
    \end{itemize}
\end{enumerate}

\paragraph{Parameter Summary}
Table~\ref{tab:hdbscan_params} summarizes the key parameters of the Hybrid Rolling HDBSCAN algorithm.\\

\begin{table}[h]
\centering
\caption{Hybrid Rolling HDBSCAN parameters.}
\label{tab:hdbscan_params}
\begin{tabular}{l|p{0.55\columnwidth}|c}
\toprule
\textbf{Parameter} & \textbf{Description} & \textbf{Default} \\
\midrule
$W_{\text{size}}$ & Sliding window size & 21 days \\
$S_{\text{step}}$ & Window step size & 3 days \\
$\tau_{\text{drift}}$ & Normalized drift threshold for contamination detection & 2.2 \\
$\tau_{\text{match}}$ & Normalized distance threshold for cluster matching & 1.0 \\
$P_{\text{drift}}$ & Percentage of drifted clusters triggering full retrain & 10\% \\
$P_{\text{dissolve}}$ & Percentage of dissolved clusters triggering full retrain & 15\% \\
$\tau_{\text{large}}$ & Large cluster size threshold & 50 \\
$\tau_{\text{unassigned}}$ & Unassigned ratio triggering retrain & 60\% \\
$M_{\text{dormant}}$ & Dormant cluster memory duration & 3 windows \\
\bottomrule
\end{tabular}
\end{table}

\textbf{CAMPAIGN IDENTIFICATION PIPELINE\\}

Initial observation of resulting clusters revealed a fundamental difference between software and gaming campaigns. Gaming campaigns are recognizable by unique executor or mod names that are semantically isolated from the surrounding vocabulary of the infection theme (e.g., "Solara," "Azxus"), while software campaigns are defined by specific product–version pairs (e.g., Photoshop\_v25.3.1.122). 

The key insight is that text embeddings capture semantic similarity but largely ignore numerical differences. Clustering alone is therefore sufficient to identify gaming campaigns, since their identifiers are semantically distinct. For software campaigns, however, embeddings treat different version numbers as nearly identical, meaning clustering alone cannot reliably separate them. Each ecosystem thus requires a tailored approach, and these structural differences drive every design choice in the pipeline, starting with how we name clusters. 

\subsection{Cluster Naming Strategies}

\subsubsection{Gaming Ecosystem (Word Frequency-Based Naming)}
For gaming-related infections (e.g., ROBLOX, GTA), clusters are named using the top-$k$ most frequent words from constituent infection themes, excluding common stopwords. This produces names such as \textit{"ROBLOX, executor, valex, free, download"}. 

\subsubsection{Software Ecosystem (Product-Level Naming)}

Software infection themes exhibit a fundamentally different structure from gaming brands infection themes. These themes contain highly specific version numbers, editions, author attributions, and product names (e.g. \textit{"Adobe Photoshop v25.3.1.122 multilingual crack by m0nkrus"}). 

As stated above, text embeddings inherently place less weight on numerical sequences than semantic content. Since text embeddings cannot reliably distinguish version numbers, clustering naturally groups infections by product rather than by specific version. We treat these clusters as product groups, not campaigns, and name them accordingly: we extract the product name from each theme and select the most frequent one across the cluster, producing names like "photoshop," "acrobat\_pro," or "premiere\_pro." Version-level differentiation is handled in the next stage of the pipeline.

\subsection{Relevance Scoring}

\subsubsection{Gaming Ecosystem - Semantic Isolation Detection}\

In the gaming ecosystem, up to half of all clusters for a given brand can be irrelevant, meaning they contain only generic terms (e.g., "free," "download," "executor") with no identifiable campaign. To filter these out, we introduce Semantic Isolation Detection (SID), a metric that identifies clusters containing a unique campaign name.

The core insight is that campaign-specific words like "valex", "kiddions" stand out from the generic vocabulary around them; they are semantically isolated. These words act as campaign identifiers: if a cluster contains one, it is relevant. If it does not, it is generic and filtered out.

To detect campaign identifiers, we check each word $w$ in a cluster name against four binary signals (Table~\ref{tab:isolation_signals}), each measuring whether the word is semantically connected to the other words in the name.

\begin{table}[h]
\centering
\caption{Semantic isolation signals for campaign identifier detection.}
\label{tab:isolation_signals}
\begin{tabular}{c|l|c|c}
\toprule
\textbf{Signal} & \textbf{Question Asked} & \textbf{YES} & \textbf{NO} \\
\midrule
$s_1$ & Does this word pair with another word? & 0  & 1  \\
$s_2$ & Is this word part of a semantic group with others? & 0 & 1 \\
$s_3$ & Does this word share a category with others? & 0 & 1 \\
$s_4$ & Is this the ONLY specific word? & 1  & 0 \\
\bottomrule
\end{tabular}
\end{table}

The isolation score for word $w$ is the sum of the four signals:

\begin{equation}
    \text{IsolationScore}(w) = \sum_{i=1}^{4} s_i(w)
\end{equation}

\noindent A word scoring 3 or higher is classified as a campaign identifier. A cluster containing at least one campaign identifier is considered relevant.  \\ 

\subsubsection{Software Ecosystem - Relevance at Subcluster Level}\

In the software ecosystem, relevance is not assessed at the product-cluster level. All product clusters are considered potentially relevant, as they merely group together themes referring to the same software product. Software campaigns are only defined after this stage, through subclustering based on version identifiers (see Section~\ref{section_software_product_to_campaign}). Relevance is therefore evaluated at the subcluster level, not during initial clustering. 

\subsection{Campaign Grouping}
Beyond individual cluster relevance, we implement campaign grouping to track related clusters across time windows as part of the same campaign.

\subsubsection{Gaming Ecosystem - Identifier Set Matching}
Gaming campaigns are grouped using Jaccard similarity on extracted campaign identifiers. Given two clusters with identifier sets $I_1$ and $I_2$, we compute Jaccard similarity:

\begin{equation}
    \text{similarity}(C_1, C_2) = \frac{|I_1 \cap I_2|}{|I_1 \cup I_2|}
\end{equation}
\noindent where $I_1$ and $I_2$ are the normalized identifier sets for clusters $C_1$ and $C_2$. Clusters with $\text{similarity} \geq 0.6$ (60\% overlap) are assigned to the same campaign.  A campaign is then seen as \textbf{actionable} if it contains at least one valid URL, if not, it is labeled as "worth watching".

\subsubsection{Software Ecosystem - Hierarchical Grouping (Product to Campaign)}
\label{section_software_product_to_campaign}
Software campaign tracking follows a two-level hierarchy that addresses the embedding limitation described above:

\paragraph{Level 1: Product Grouping}

The first step groups clusters by product name rather than defining campaigns. All clusters sharing the same product name, for example, "photoshop", are merged into a single product group, regardless of version. These product groups simply collect related infections before the next step splits them by version into actual campaigns. The one exception is author-based grouping: if an author is detected (e.g., "ratiborus," "m0nkrus"), the cluster is assigned directly to an author group, which takes priority over product grouping.

\paragraph{Level 2: Campaign Extraction (Version-Based)}

Within each product group, campaigns are defined by version number. We parse version strings from infection themes and group infections sharing the same product–version pair into a single campaign. For example, the photoshop product group may yield three distinct campaigns: photoshop\_v25.3.1.122, photoshop\_v25.1.0.0, and photoshop\_v24.7.0. Infections with no version identifier are considered irrelevant and excluded. 

In the same fashion as the gaming campaigns, a software campaign is considered actionable if
it contains at least one valid URL, if not, it is labeled as "worth watching".

\begin{figure*}[t]
\centering
\caption{Ecosystem-specific campaign discovery pipelines. Gaming campaigns are identified directly at the cluster level, while software campaigns require a hierarchical product-to-version extraction due to embedding limitations on numerical data.}
\label{fig:pipeline_comparison}

\begin{tikzpicture}[
    node distance=1.0cm,
    every node/.style={
        draw,
        rectangle,
        rounded corners,
        align=center,
        minimum height=0.7cm,
        minimum width=2.6cm,
        font=\footnotesize
    },
    arrow/.style={->, line width=0.5pt, >={Latex[length=2mm, width=1.5mm]}}
]

\node (g1) {Clusters};
\node (g2) [right=of g1] {Relevance Filtering\\(SID)};
\node (g3) [right=of g2] {Campaign Grouping\\(Identifier Similarity)};
\node (g4) [right=of g3] {Actionable Campaigns (If URL)};

\draw[arrow] (g1) -- (g2);
\draw[arrow] (g2) -- (g3);
\draw[arrow] (g3) -- (g4);

\node[draw=none, above=0.1cm of g1] {\textbf{Gaming}};

\node (s1) [below=1.8cm of g1] {Clusters};
\node (s2) [right=of s1] {Product Grouping\\(Product-Level Aggregation)};
\node (s3) [right=of s2] {Version-Based\\Campaign Extraction};
\node (s4) [right=of s3] {Relevance Assessment\\(version presence)};
\node (s5) [right=of s4] {Actionable Campaigns (If URL)};

\draw[arrow] (s1) -- (s2);
\draw[arrow] (s2) -- (s3);
\draw[arrow] (s3) -- (s4);
\draw[arrow] (s4) -- (s5);

\node[draw=none, above=0.1cm of s1] {\textbf{Software}};

\end{tikzpicture}
\end{figure*}

Figure~\ref{fig:pipeline_comparison} illustrates both pipelines, and Table~\ref{tab:category_comparison} summarizes the key methodological differences between gaming and software ecosystems.

\begin{table*}[t]
\centering
\caption{Comparison of ecosystem-specific processing strategies.}
\label{tab:category_comparison}
\renewcommand{\arraystretch}{1.3}
\begin{tabular}{l|c|c}
\toprule
\textbf{Aspect} & \textbf{Gaming Ecosystem} & \textbf{Software Ecosystem} \\
\midrule
First clustering & Campaign identifier (name-based) & Product name \\
Cluster naming strategy & Top-$k$ frequent tokens & Product extraction \\
Are clusters campaigns? & Often yes & No \\
Campaign definition & Identifier-based grouping & Product + version \\
Relevance assessment level & Cluster & Campaign \\
Actionability criterion & Campaign with $\geq 1$ URL & Campaign with $\geq 1$ URL \\
\bottomrule
\end{tabular}
\end{table*}

\subsection{Campaign Identification Assessment Framework}

To evaluate the effectiveness of our campaign discovery pipeline, we establish a ground truth validation framework and define metrics that capture both the accuracy and quality of campaign identification. Framework evaluation is conducted on four high-volume brands, two per ecosystem (KMS and Adobe for software, ROBLOX and GTA for gaming), which contain sufficient data density and distinct campaigns to support meaningful precision and recall analysis. The remaining brands did not exhibit enough distinct, observable campaigns to establish reliable ground truth and were therefore excluded from quantitative evaluation.

\subsubsection{Ground Truth Construction}

We manually constructed ground truth datasets for each brand by analyzing infection themes for specificity and frequency. Campaigns were stratified by prevalence:

\begin{itemize}
    \item \textbf{Campaign Sizes:} Very large campaigns contain over 100 infections, Large campaigns contain at least 50 infections, medium campaigns at least 25, small campaigns at least 10 and emerging campaigns at least 5.
\end{itemize}

Each ground truth campaign is represented by one or more \textit{identifiers}, unique tokens that unambiguously identify the campaign. Identifiers are derived from game executor and mod menu names for gaming brands (e.g. "xeno", "solara" for ROBLOX; "kiddions" for GTA), repackager, and/or product-version combinations for software brands.

Each ground truth identifier represents a distinct campaign that should ideally be discovered as a separate entity by our framework.

\subsubsection{Matching Criteria}

Matching criteria differ by ecosystem:

\begin{itemize}
    \item \textbf{Gaming:} To be considered a match, a framework-identified campaign's extracted identifiers must match at least one of the ground truth campaign's identifiers.
    \item \textbf{Software:} To be considered a match, a framework-identified campaign must match the exact product and version of the ground truth campaign.
\end{itemize}

We define four metrics to assess campaign identification accuracy. Recall and precision apply to both gaming and software ecosystems. Over-grouping and under-grouping are specific to the gaming ecosystem, where Jaccard similarity-based clustering can merge or fragment campaigns in ways that version-specific software matching does not.

\paragraph{Recall} The proportion of ground truth campaigns successfully identified by the framework.

\paragraph{Precision} The proportion of framework-identified campaigns that correctly correspond to a ground truth campaign.

\paragraph{Over-grouping (Gaming Only)} The proportion of framework-identified campaigns that erroneously merge multiple ground truth campaigns into one. Over-grouping is a critical failure mode: when distinct campaigns are merged, one becomes effectively invisible, impeding targeted remediation.

\paragraph{Under-grouping (Gaming Only)} Occurs when a single ground truth campaign is split across multiple framework-identified campaigns. While suboptimal, under-grouping preserves visibility of all campaigns and does not hinder remediation. We therefore report it as a descriptive statistic rather than a penalizing metric.

\section{RESULTS}

\subsection{Clustering Performance}

To evaluate the clustering quality of the Hybrid Rolling HDBSCAN framework, we focus on the two most prevalent gaming brands (ROBLOX and GTA) and two of the most prevalent software brands (KMS and Adobe), with Microsoft reported in Table~\ref{tab:clustering_metrics_combined} for reference.

Within gaming-related infections, GTA ranks first with 2,623 infection themes, while ROBLOX ranks second with 2,371 themes. In the software domain, KMS is the most prevalent brand overall with 3,049 infection themes and Adobe is not far behind with 2,301 themes (see Table ~\ref{tab:software_distribution}).

\subsubsection{Gaming Brands}
Clustering quality metrics for gaming brands are consistent across both datasets as presented in Table ~\ref{tab:clustering_metrics_combined}. The average silhouette score across windows remains around 0.4 ($\overline{S}_{\text{ROBLOX}} = 0.403$, $\overline{S}_{\text{GTA}} = 0.386$), indicating moderate cluster separation in a high-noise setting. Correspondingly, the Davies--Bouldin Index remains just above 1.0 ($\overline{DBI}_{\text{ROBLOX}} = 1.030$, $\overline{DBI}_{\text{GTA}} = 1.009$), reflecting the expected overlap induced by generic terms and shared vocabulary across campaigns.

Despite this noise, gaming clusters exhibit strong temporal stability, with persistence rates exceeding 80\% and average lifespans of approximately four windows. This suggests that, although clusters are not perfectly separated geometrically, they remain coherent enough over time to support campaign tracking.

\subsubsection{Software Brands}
Software brands exhibit notably stronger clustering performance. Average silhouette scores exceed 0.54 for both KMS and Adobe ($\overline{S}_{\text{KMS}} = 0.583$, $\overline{S}_{\text{Adobe}} = 0.542$), while the Davies--Bouldin Index remains between 0.70 and 0.81 ($\overline{DBI}_{\text{KMS}} = 0.711$, $\overline{DBI}_{\text{Adobe}} = 0.802$), indicating tighter and better-separated clusters.

This improvement is largely attributable to the structured nature of software infection themes and their repetitiveness, which reduces semantic ambiguity and limits overlap between unrelated campaigns.

\begin{table*}[t]
\centering
\caption{Clustering quality metrics for Hybrid Rolling HDBSCAN across gaming and software ecosystems.}
\label{tab:clustering_metrics_combined}
\begin{tabular}{l|cc|ccc}
\toprule
\textbf{Metric} &
\multicolumn{2}{c|}{\textbf{Gaming}} &
\multicolumn{3}{c}{\textbf{Software}} \\
\cmidrule(lr){2-3} \cmidrule(lr){4-6}
 & \textbf{ROBLOX} & \textbf{GTA} & \textbf{KMS} & \textbf{Microsoft} & \textbf{Adobe}\\
\midrule

\multicolumn{5}{l}{\textit{Internal Validity}} \\
\midrule
Silhouette Score &
0.403 $\pm$ 0.018 &
0.386 $\pm$ 0.018 &
0.583 $\pm$ 0.014 &
0.635 $\pm$ 0.015 &
0.542 $\pm$ 0.030\\

Davies--Bouldin Index &
1.030 $\pm$ 0.045 &
1.009 $\pm$ 0.039 &
0.711 $\pm$ 0.030 &
0.675 $\pm$ 0.041 &
0.802 $\pm$ 0.032\\

\midrule
\multicolumn{5}{l}{\textit{Temporal Stability}} \\
\midrule
Persistence Rate &
82.4\% &
81.9\% &
87.7\% &
79.4\% &
75.4\% \\

Mean Lifespan (windows) &
4.2 $\pm$ 2.6 &
4.4 $\pm$ 2.9 &
5.1 $\pm$ 2.8 &
3.7 $\pm$ 2.2 &
3.0 $\pm$ 1.7 \\

\midrule
\multicolumn{5}{l}{\textit{Operational}} \\
\midrule
Noise Ratio &
58.9\% $\pm$ 4.2\% &
54.4\% $\pm$ 3.8\% &
45.0\% $\pm$ 3.0\% &
49.3\% $\pm$ 5.0\% &
61.0\% $\pm$ 5.8\%\\

Mean Cluster Count &
18.8 $\pm$ 3.9 &
25.3 $\pm$ 8.3 &
30.7 $\pm$ 6.5 &
21.8 $\pm$ 7.3 &
18.4 $\pm$ 2.6\\

Retrain Frequency &
88.9\% &
88.9\% &
55.6\% &
87.5\% &
87.5\% \\

\bottomrule
\end{tabular}
\end{table*}

\subsection{Relevance Filtering Descriptive Results}

Table~\ref{tab:relevance_statistics} summarizes the impact of the relevance detection systems applied to gaming. The Semantic Isolation Detection classified 30 (63.8\%) ROBLOX clusters as relevant, filtering out 17 clusters (36.2\%) that contained only generic terms such as "free", "download", or "executor". GTA exhibited a higher filtering rate of 46.2\%, with 14 of 26 clusters classified as relevant.

This disparity reflects differences in infection theme characteristics between both games. GTA themes tend to contain more generic descriptors and fewer unique campaign identifiers compared to ROBLOX, resulting in a higher proportion of clusters lacking distinctive signals. Examples of filtered clusters include  "GTA, san, andreas, grand, download" or "fivem, gta, hack, aimbot, 2025", clusters that provide no actionable campaign information.

Overall, gaming brands required substantial filtering (36--46\%) to isolate clusters with campaign-specific identifiers.

\begin{table}[t]
\centering
\caption{Relevance Filtering Statistics.}
\label{tab:relevance_statistics}
\begin{tabular}{l|cc}
\toprule
\textbf{Metric} &
\multicolumn{2}{c}{\textbf{Gaming}} \\
\cmidrule(lr){2-3} 
 & \textbf{ROBLOX} & \textbf{GTA} \\
\midrule
Total Clusters & 47 & 26 \\
Relevant  & 30 & 14  \\
Filtered  & 17 & 12 \\
Filtering Rate & 36.2\% & 46.2\% \\
\bottomrule
\end{tabular}
\end{table}

\subsection{Campaign Identification Accuracy}

\textbf{GAMING ECOSYSTEM}

\subsubsection{ROBLOX}

\paragraph{Ground Truth}
Manual analysis identified 25 distinct campaigns:
\begin{itemize}
\item \textbf{Large} (8 campaigns): Xeno (282 infections), Swift (138), Solara (134), Valex\_Byfron (106), Delta (97),   Grow a Garden (77), Luna (73), Blox Fruit (62)
\item \textbf{Medium} (4 campaigns): Lua (39), Jailbreak (31), Nezur (31), Velocity (30) 
\item \textbf{Small} (10 campaigns): Rivals (23), Fluxus\_Quantum (22), Synapse X (22), FPS Boost (18), Arsenal (15), Speed Hub X (14), Azrix (13), Electron (11), Arcane (11), Fisch (10)
\item \textbf{Emerging} (3 campaigns): Codex (8), Voidstrap (7), Hood Zapped (6)
\end{itemize}

\paragraph{Results}
The framework identified 19 campaigns. Cross-referencing with ground truth yields the results shown in Table~\ref{tab:ROBLOX_accuracy}. Overall, our framework identified 68\% (17/25) of ground truth campaigns, with errors concentrated in small and emerging campaigns, plus one medium campaign (Velocity) hidden through over-grouping. Our framework achieved a precision of 94.74\% (18/19), with only one over-grouped campaign (Velocity\_swift) reducing precision. The under-grouped campaign did not affect precision, as all clusters still corresponded to valid ground truth campaigns.

\begin{table}[h]
\centering
\caption{ROBLOX campaign identification results}
\label{tab:ROBLOX_accuracy}
\begin{tabular}{lcp{4cm}}
\toprule
\textbf{Outcome} & \textbf{Count} & \textbf{Campaigns} \\
\midrule
Correctly identified & 17 & Xeno, Solara, Swift, Nezur, Grow a Garden, Azrix, \\
& & Luna, Jailbreak, Arcane, Lua, Blox Fruit,\\ & & Valex\_Byfron, Electron, Delta, Fluxus\_Quantum,\\  & & Hood\_Zapped, FPS\_Boost\\
Under-grouped & 1 & Delta\_ios\\
Over-grouped & 1 & Velocity\_swift \\
Missed & 7 & Arsenal, Voidstrap, Rivals, Synapse X, Speed Hub X, Codex, Fisch \\
\bottomrule
\end{tabular}
\end{table}

\subsubsection{GTA}

\paragraph{Ground Truth}
Manual analysis identified 18 distinct campaigns:
\begin{itemize}
\item \textbf{Large} (2 campaigns): Kiddions (709 infections), Yim (119)
\item \textbf{Medium} (5 campaigns): FitGirl (35), Lua (34), Spoofer (30), FPS Boost (29), Rampage (25)
\item \textbf{Small} (9 campaigns): QuantV (19), Eulen (17), Crash (16), Rune (16), Cherax (14), Connection Fix (13), TZ Project (12), Dodi (11), Nenyoo (10)
\item \textbf{Emerging} (2 campaigns): CPU Usage (6), Gamingbeast (5)
\end{itemize}

\paragraph{Results}
The framework identified 16 campaigns. Table~\ref{tab:gta_accuracy} summarizes the results. For GTA, our framework correctly identifies 83.33\% (15/18) of ground truth campaigns, with no over-grouping observed. Precision for GTA is 100\%, as all 16 identified campaigns correspond to ground truth campaigns. Under-grouping, which split one campaign (Yim) into multiple clusters, does not reduce precision since no campaigns were false positives.

\begin{table}[h]
\centering
\caption{GTA campaign identification results}
\label{tab:gta_accuracy}
\begin{tabular}{lcp{4cm}}
\toprule
\textbf{Outcome} & \textbf{Count} & \textbf{Campaigns} \\
\midrule
Correctly identified & 15 & Kiddions, FitGirl, Rampage, Connection Fix, Crash, \\
& & Gamingbeast, QuantV, Nenyoo, CPU Usage, Lua, Cherax, Rune, Spoofer, Yim, FPS Boost \\
Under-grouped & 1 & money\_yim \\
Over-grouped & 0 & --- \\
Missed & 3 & Dodi, TZ Project, Eulen \\
\bottomrule
\end{tabular}
\end{table}

\subsubsection{Error Analysis and Failure Modes}

\paragraph{Missed Campaigns}
Across both ROBLOX and GTA, missed campaigns are exclusively small-scale or emerging, suggesting our gaming framework performs better on more populated campaigns rather than identifying smaller-scale ones. Interestingly enough, our framework successfully identified both emerging activity campaigns for GTA and one of three for ROBLOX. While our framework can identify smaller campaigns and emerging campaigns, the performance is less reliable: combining both ROBLOX and GTA, our gaming framework was able to identify 58.33\% (14/24) of small and emerging activity campaigns. 

\paragraph{Under-grouping}
Under-grouping occurs when a single ground truth campaign is fragmented into multiple framework campaigns. In gaming brands, GTA shows 5.56\% under-grouping (1/18 GT campaigns were under-grouped) and ROBLOX shows a 4\% under-grouping (1/25 GT campaigns were under-grouped). Fragmentation is driven by thematic variation within infection themes, such as:
\begin{itemize}
\item platform-specific variants (e.g. Delta),
\item feature-focused descriptions (e.g. Yim money drops),
\end{itemize}
Under-grouping preserves campaign visibility but increases analyst workload, as related campaigns must be manually consolidated.

\paragraph{Over-grouping}
Over-grouping represents the most severe failure mode, as it can hide campaigns within merged campaigns. Only one instance occurred in ROBLOX: Velocity\_swift merged the Swift and Velocity campaigns. Swift also appears as a standalone campaign, but Velocity is only present in the merged campaign, making it effectively hidden. No over-grouping was observed in GTA.\\

\textbf{SOFTWARE ECOSYSTEM}

In the software ecosystem, campaigns are defined by exact product, variant, and version, resulting in binary matching outcomes. Over and under-grouping do not occur under this definition. So, recall reflects the framework’s ability to recover known campaigns, while precision captures output noise, including early-stage clusters and naming mismatches.

\subsubsection{Adobe}

\paragraph{Ground Truth}
Manual analysis identified 36 distinct campaigns:
\begin{itemize}
\item \textbf{Author-based} (1): M0nkrus
\item \textbf{Product-based} (35)
\end{itemize}

These campaigns break down into 12 emerging ($x\geq5$ infections), 17 small ($x\geq10$ infections), 5 medium ($x\geq25$ infections), and 2 large ($x\geq50$ infections). 

\paragraph{Results}
The framework identified 32 campaigns: 1 author-based and 31 product-version-based. Overall, 32 of the 36 ground truth campaigns were correctly identified, yielding a recall of 88.89\% (32/36). 

All campaigns with more than 10 infections were identified (Table~\ref{tab:adobe_accuracy}). Our framework missed one small ground truth campaign (10 infections) and three emerging campaigns (5, 5 and 6 infections, respectively).

\begin{table}[h]
\centering
\caption{Adobe campaign identification results}
\label{tab:adobe_accuracy}
\begin{tabular}{lc}
\toprule
\textbf{Outcome} & \textbf{Count} \\
\midrule
Correctly identified & 32 \\
\quad Author-based & 1 \\
\quad Product-version-based & 31 \\
Missed & 4 \\
\quad Product misnaming error & 0 \\
\bottomrule
\end{tabular}
\end{table}

\subsubsection{KMS}

\paragraph{Ground Truth}
Manual analysis identified 33 distinct campaigns:
\begin{itemize}
\item \textbf{Author-based} (3): Ratiborus
\item \textbf{Product-based} (30)
\end{itemize}

These campaigns break down into 9 emerging ($x\geq5$), 12 small ($x\geq10$), 6 medium ($x\geq25$), 3 large ($x\geq50$), and 3 very large ($x\geq100$). 

\paragraph{Results}
The framework identified 27 campaigns: 3 author-based and 24 product-version-based. Of the 33 ground truth campaigns, 26 were correctly identified, resulting in a recall of 78.78\% (26/33).

All campaigns with at least 10 infections were identified, yielding a 100\% recall for small, medium, large and very large campaigns (Table~\ref{tab:kms_accuracy}). 
All missed campaigns fall below this threshold (1: 8 infections, 1: 7 infections, 2: 6 infections, 2: 5 infections). All three author-based campaigns were correctly identified by our framework.

\begin{table}[h]
\centering
\caption{KMS campaign identification results}
\label{tab:kms_accuracy}
\begin{tabular}{lc}
\toprule
\textbf{Outcome} & \textbf{Count} \\
\midrule
Correctly identified & 26 \\
\quad Author-based & 3 \\
\quad Product-version-based & 23 \\
Missed & 6 \\
\quad Extra & 1 \\
\bottomrule
\end{tabular}
\end{table}

\subsubsection{Analysis of Missed and Additional Campaigns}

\paragraph{Missed Campaigns}
Across both KMS and Adobe, missed campaigns are consistently low-infection clusters, all at or below 10 infections. This indicates that the framework's software approach reliably detects established campaigns but is less consistent for emerging ones. For campaigns with 10 or more infections, detection reaches 95.83\% (23/24) for Adobe and 100\% for KMS.

\paragraph{Additional Identified Campaigns}
The framework also identified one campaign that is not present in the ground truth for KMS. This campaign corresponds to early-stage emerging activity; as it did not meet the five-theme threshold, it was not included in our ground-truth. This cluster may indicate emerging activity and could be valuable for early monitoring. 

\subsubsection{Cross-Ecosystem Accuracy Summary}

Table~\ref{tab:accuracy_summary} presents campaign identification accuracy across all ecosystems. 

\begin{table}[h]
\centering
\caption{Campaign identification accuracy across gaming and software ecosystems.}
\label{tab:accuracy_summary}
\setlength{\tabcolsep}{7pt}
\renewcommand{\arraystretch}{1.3}
\begin{tabular}{lcccc}
\toprule
 & \multicolumn{2}{c}{\textbf{Gaming}} & \multicolumn{2}{c}{\textbf{Software}} \\
\cmidrule(lr){2-3} \cmidrule(lr){4-5}
\textbf{Metric} & \textbf{ROBLOX} & \textbf{GTA} & \textbf{KMS} & \textbf{Adobe} \\
\midrule
\multicolumn{5}{l}{\textit{Campaign Counts}} \\[2pt]
\quad Ground truth campaigns & 25 & 18 & 33 & 36 \\
\quad Identified campaigns & 19 & 16 & 27 & 32 \\
\quad Correctly identified & 17 & 15 & 26 & 32 \\
\midrule
\multicolumn{5}{l}{\textit{Core Metrics}} \\[2pt]
\quad Recall & 68\% & 83.33\% & 78.78\% & 88.89\% \\
\quad Precision & 94.74\% & 100\% & 96.30\% & 100\% \\
\midrule
\multicolumn{5}{l}{\textit{Error Analysis}} \\[2pt]
\quad Missed campaigns & 7 & 3 & 6 & 4 \\
\quad Under-grouped & 1 & 1 & --- & --- \\
\quad Over-grouped & 1 & 0 & --- & --- \\
\quad Extra & 0 & 0 & 1 & 0 \\
\bottomrule
\end{tabular}
\end{table}

\subsubsection{Campaign Population Completeness}
\label{sec: completeness}
Beyond identifying campaigns, we evaluate how completely our framework captures each campaign's population. \textit{Completeness} measures the proportion of a campaign's ground truth infections that appear in the corresponding framework cluster. Values exceeding 100\% indicate \textit{overpopulation}, the inclusion of infections not belonging to the ground truth campaign.

\paragraph{Software Ecosystem}
Software brands achieved a mean completeness of 94.56\%, indicating that our framework successfully clustered nearly all infections belonging to each campaign's population. This high completeness reflects the homogeneity of software infection themes, which typically follow consistent patterns (product + version + author). Many campaigns contain near-identical infection themes repeated across dozens of infections, making semantic clustering easier.

However, 7 of 59 software campaigns identified (11.86\%) exhibited overpopulation, with excess infections ranging from 1 to 7 per campaign. These cases represent minor contamination that does not substantially compromise campaign integrity.

\paragraph{Gaming Ecosystem}
Gaming brands achieved a lower mean completeness of 71.13\%. While 80\% of identified campaigns captured over half of their ground truth population, the remaining 20\% achieved completeness rates below 50\%. This reduction stems from the semantic heterogeneity of gaming infection themes.  Infection themes within a single campaign can vary substantially, causing some variants to fall outside semantic similarity thresholds.

For example, the Xeno campaign contained the following ground truth themes:
\begin{itemize}
    \item \texttt{"Xeno Executor v1.2.3"}
    \item \texttt{"XENO PC V1.1.8, CRYPTIC PC (PATCH) - free executor tools for Windows."}
    \item \texttt{"ROBLOX executor version Xeno v1.2.35"}
\end{itemize}
When intermediate or bridging themes are absent, some variants may not be clustered together, or remain unassigned, resulting in partial population recovery, reducing population completeness.

\paragraph{Overpopulation in Gaming}
Gaming brands exhibited more severe overpopulation. Of 35 gaming campaigns, 5 cases occurred, with excess infections ranging from 1 to 68. The most significant case involved the Electron campaign: our framework correctly identified all 11 ground truth infections but additionally clustered 68 unrelated infections. Manual inspection revealed these extra infections contained generic themes such as:
\begin{itemize}
    \item \texttt{"ROBLOX exploit executor"}
    \item \texttt{"ROBLOX tool executor"}
\end{itemize}
These themes share semantic similarity with legitimate Electron infections (e.g. \texttt{"ROBLOX electron executor"}), causing the clustering algorithm to merge them. This behavior is representative of most overpopulation cases in gaming. Infection themes frequently contain generic descriptors (e.g., executor, tool, exploit) that appear across multiple campaigns, increasing the likelihood of unintended cluster expansion when semantic similarity is high. This failure mode highlights a limitation of semantic clustering when campaigns share vocabulary with generic ecosystem-wide terminology.

Table~\ref{tab:completeness_summary} summarizes completeness metrics across ecosystems.

\begin{table}[h]
\centering
\caption{Campaign population completeness across ecosystems.}
\label{tab:completeness_summary}
\begin{tabular}{lcc}
\toprule
\textbf{Metric} & \textbf{Gaming} & \textbf{Software} \\
\midrule
Mean completeness & 71.13\% & 94.56\% \\
Overpopulation cases & 5/35 (14.29\%) & 7/59 (11.86\%) \\
Max overpopulation & +68 infections & +7 infections \\
\bottomrule
\end{tabular}
\end{table}

Overall, campaign population completeness closely reflects the semantic cleanliness of infection themes. Software campaigns benefit from consistent, product-centric naming conventions, enabling near-perfect population recovery. Gaming campaigns, by contrast, exhibit high semantic variability, leading to both under-coverage and occasional overpopulation.

\subsection{Structural and Temporal Characterization of Campaigns}

This section analyzes the structural and temporal characteristics of campaigns as identified by our framework. Campaign analysis includes all top ten brands from Table~\ref{tab:software_distribution} to capture the full breadth of observed activity. While framework accuracy was validated only on the four high-volume brands due to data density requirements, all brands are included in aggregate analyses (software vs. gaming) to maximize campaign coverage for comparative insights. 

The study of campaigns therefore covers just over a third (34.32\%) of the initial sample, encompassing 18,496 infections across ten brands: five software brands (Adobe, KMS, Internet Download Manager, Microsoft, and Windows) and five gaming brands (ROBLOX, GTA, Valorant, Fortnite, and Counter Strike). Within this subset, the framework identified 177 campaigns in total, including 128 software campaigns and 49 gaming campaigns. With 99\% of records dated before July 25th, campaigns were identified from infections between June 7th and July 25th, 2025. Therefore, all temporal measures below are bounded by this 47-day window (see Section~\ref{sec:limitations}).

\subsubsection{URL Actionability and Distribution Patterns}

A campaign is considered \textit{actionable} if it contains at least one URL, as URLs provide concrete pivot points for investigation, takedown requests, or threat attribution.

\paragraph{Software Campaigns}

Of the 128 software campaigns, 84 (65.6\%) were actionable, i.e. referenced at least one URL, averaging 2.87 URLs per campaign (median: 2).

Torrent indexes dominate (Table~\ref{tab:software-domains}): 51 campaigns (39.8\%) reference at least one. Most of them link to the same site, The Pirate Bay (35 campaigns, 27.3\%), followed by 1337x (12 campaigns, 9.4\%). The Pirate Bay is reached through 20 different mirror domains (e.g., \texttt{tpb.party}, \texttt{thepiratebay0.org}). General hosting services are rarely used (GitHub: 5, Amazon S3: 4, MEGA: 3).

\begin{table}[h]
\centering
\begin{tabular}{lrr}
\toprule
\textbf{Platform} & \textbf{Campaigns} & \textbf{\% of Software} \\
\midrule
Torrent indexes      & 51 & 39.8\% \\
\quad The Pirate Bay   & 35 & 27.3\% \\
\quad 1337x           & 12 & 9.4\% \\
uTorrent Web\textsuperscript{a}           & 10 & 7.8\% \\
GitHub                                    & 5  & 3.9\% \\
Amazon S3                                 & 4  & 3.1\% \\
MEGA                                      & 3  & 2.3\% \\
\midrule
\textit{Standalone websites}\textsuperscript{b} & 39 & 30.5\% \\
\bottomrule
\end{tabular}
\caption{Most common URL platforms in software campaigns ($n=128$). A campaign can reference several platforms.}
\label{tab:software-domains}

{\footnotesize 
\textsuperscript{a}\texttt{utweb.rainberrytv.com} / \texttt{btweb.rainberrytv.com}, the uTorrent and BitTorrent Web client interface.\\
\textsuperscript{b}Any domain outside the platforms above (43 distinct registered domains).\par}
\end{table}

Standalone websites appear in 39 campaigns (30.5\%), mostly cracked-software aggregators such as \texttt{crackingpatching.com} (6). They are rarely shared: only 6 of 43 such domains appear in more than one campaign. Software campaigns thus rely on established piracy infrastructure rather than dedicated sites, unlike gaming campaigns.

\paragraph{Gaming Campaigns}

Gaming campaigns contain more URLs than software campaigns in our sample: 44 of the 49 gaming campaigns (89.8\%) contain at least one URL, compared to 65.6\% of software campaigns. Mainstream platforms are widely shared across campaigns (Table~\ref{tab:gaming-domains}): GitHub in 22 campaigns (44.9\%), MediaFire and YouTube in 15 each (30.6\%), and Discord and Google Sites in 7 each (14.3\%).

However, these platforms rarely stand alone. 41 of the 44 campaigns with URLs (93.2\%) also point to standalone websites outside mainstream platforms, and only three rely exclusively on them. These sites are numerous and mostly campaign-specific. Of 229 distinct standalone domains, only 30 appear in more than one campaign, with \texttt{dropcheats.net} (8 campaigns) and \texttt{lynxcheats.run}, \texttt{endorphin.live} and \texttt{sakpot.com} (5 each) being the most reused. Many of these sites adopt names that mimic the tool being impersonated (e.g., \texttt{xenoexecutor.live}, \texttt{kiddionsmodmenu.com}), which makes the lure look more legitimate.

These channels are typically combined inside gaming campaigns: 30 of the 44 campaigns span at least two types of sites. Every campaign linking to YouTube or Discord (18) also references a standalone site, and 13 of them additionally link to a file-sharing or code-hosting service. This suggests that YouTube videos and Discord servers are used to promote the tool and send victims to the sites that host the payload.

\begin{table}[h]
\centering
\begin{tabular}{lrr}
\toprule
\textbf{Platform} & \textbf{Campaigns} & \textbf{\% of Gaming} \\
\midrule
GitHub       & 22 & 44.9\% \\
MediaFire                        & 15 & 30.6\% \\
YouTube                          & 15 & 30.6\% \\
Discord      & 7  & 14.3\% \\
Google Sites                     & 7  & 14.3\% \\
\midrule
\textit{Standalone websites}\textsuperscript{a} & 41 & 83.7\% \\
\bottomrule
\end{tabular}
\caption{Most common URL platforms in gaming campaigns ($n=49$). A campaign can reference several platforms.}
\label{tab:gaming-domains}

{\footnotesize \textsuperscript{a}Any domain outside the platforms above (229 distinct domains).\par}
\end{table}

\subsubsection{Campaign Longevity}
\begin{figure}[h] 
\centering 
\includegraphics[width=\columnwidth]{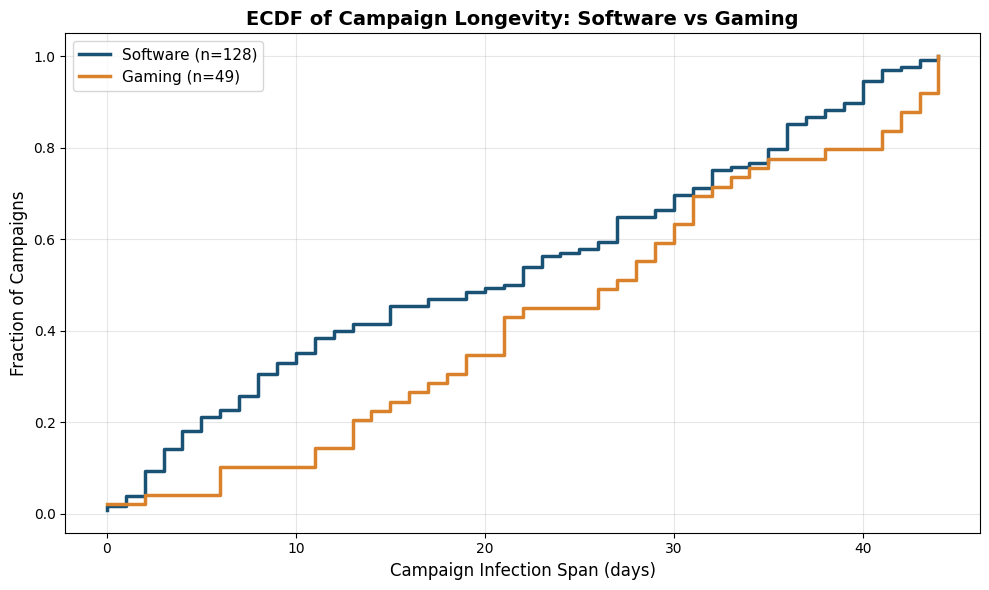}
\caption{ECDF (Empirical Cumulative Distribution Function) plot of Campaign Longevity.}
\label{fig:longevity}
\end{figure}

Figure~\ref{fig:longevity} presents the empirical cumulative distribution function (ECDF) of campaign longevity, measured as the span between first and last observed infection, aggregated by software and gaming ecosystems.

Software campaigns are shorter-lived than gaming campaigns. Nearly 50\% of software campaigns end within 20 days, compared to 35\% for gaming campaigns. The median lifespan for software campaigns is 21.5 days, while gaming campaigns persist longer with a median of 27.0 days.

This difference likely reflects operational dynamics. Software campaigns are often tied to specific versions or cracks that become obsolete as updates and patches are released, leading to faster decay. Gaming-related tools and add-ons tend to remain functional for longer periods, allowing campaigns to persist with less frequent modification.

Both ecosystems exhibit long tails, with maximum lifespans reaching 44 days, close to the length of the observation window (47 days), indicating a subset of resilient campaigns with sustained distribution. Across all campaigns, the median lifespan is 22 days, defining a practical intervention window before hitting the natural campaign decline. Overall, software campaigns decay faster, while gaming campaigns demonstrate greater persistence and longer operational windows.

\subsubsection{Temporal Dynamics of Campaign Infections}

We examine two complementary dimensions of campaign propagation: (1) how infection intensity relates to campaign longevity, and (2) how infections accumulate over time. Together, these analyses reveal different propagation patterns between software and gaming campaigns.

\paragraph{Longevity vs Intensity}

\begin{figure*}[h] 
\centering 
\includegraphics[width=\textwidth]{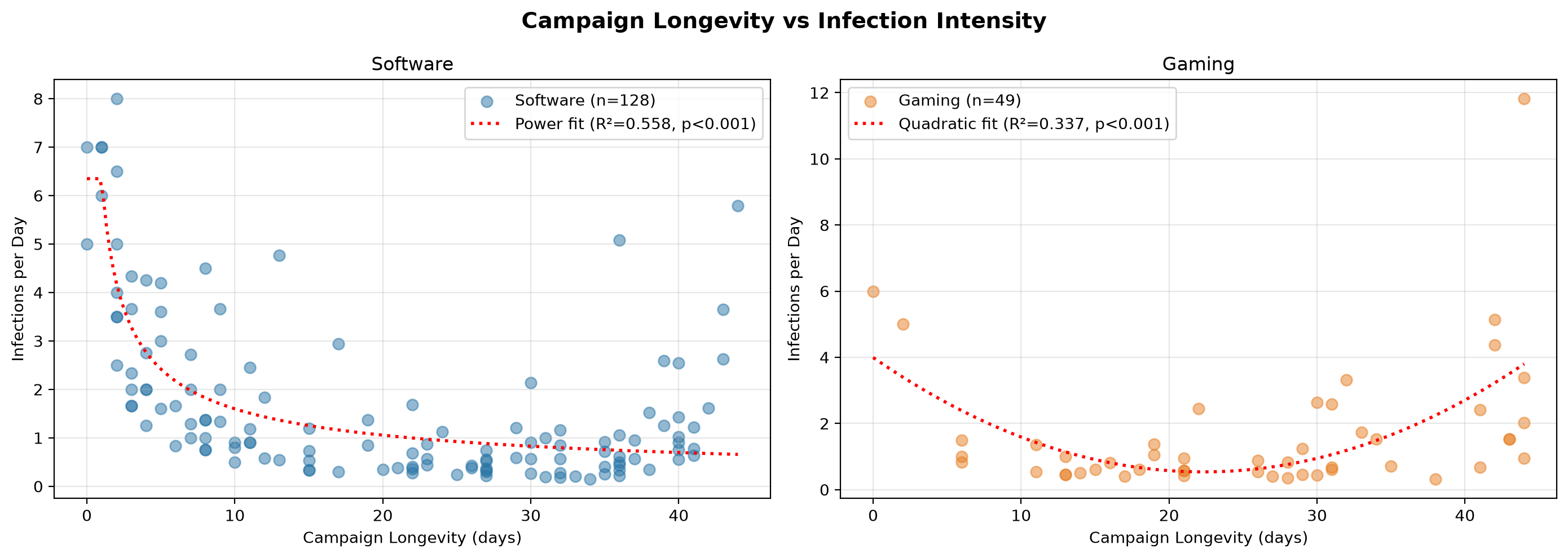}
\caption{Relation between Campaign Longevity and Infection Intensity.}
\label{fig:longevity_intensity}
\end{figure*}

Figure~\ref{fig:longevity_intensity} shows the relationship between campaign longevity and average daily infection intensity. Infection intensity is defined as the number of observed infections divided by the campaign's infection span (days between its first and last observed infection, with a minimum of one day).

Software campaigns are strongly front-loaded: short campaigns (0-7 days) achieve the highest infection rates (mean 3.48/day), declining sharply to 1.61/day for campaigns lasting 8-14 days and reaching a minimum of 0.66/day for campaigns in the 15-28 day range (Table~\ref{tab:longevity_intensity_summary}). This negative relationship is statistically significant under both Pearson ($r=-0.467$, $p<0.001$) and Spearman ($\rho=-0.489$, $p<0.001$) correlations. Model comparison by AIC favors a power-law decay ($R^2=0.558$, $p<0.001$; Table~\ref{tab:longevity_intensity_models}), with intensity falling approximately as $t^{-0.60}$: most of the decline occurs within the first two weeks, after which intensity stabilizes at a low level.

Because intensity is computed as infections over span, short-lived campaigns mechanically yield higher intensities; the accumulation analysis below (Section~\ref{sec:accumulation}), which is normalized per campaign, provides an independent confirmation of front-loading.

Long-running software campaigns (29+ days) show a higher mean intensity (1.15/day) than mid-length ones, although the preferred model does not support a systematic rebound. This uplift is driven by a small number of long-lived, high-intensity campaigns and may reflect a survivorship effect: campaigns that persist beyond the initial decay phase may represent established versions that have accumulated visibility through search indexing and community recommendation, shifting from release-driven demand to search-driven discovery.

Gaming campaigns show no significant monotonic relationship between longevity and intensity across all campaigns (Pearson: $r=0.204$, $p=0.159$; Spearman: $\rho=0.277$, $p=0.054$). A quadratic model fits best by AIC ($R^2=0.337$, $p<0.001$), but this curvature rests on the five campaigns lasting under a week (mean 2.87/day, SD 2.44), two of which exceed 5 infections/day. Excluding them, the quadratic no longer outperforms monotonic models: a monotonic power model achieves a lower AIC (49.88 vs.\ 50.20), and the quadratic improves on a linear fit by less than 2 AIC units (50.20 vs.\ 52.14). Intensity instead increases significantly with longevity ($r=0.471$, $p=0.001$; $\rho=0.510$, $p<0.001$).

Gaming campaigns lasting more than a week therefore show the opposite of software: intensity increases with longevity, and long-running campaigns reach 2.28 infections/day, almost twice the rate of long-running software campaigns. The modest explanatory power of all models suggests that gaming intensity is driven largely by factors beyond longevity, such as community-driven popularity or game update cycles.

\begin{table}[h]
\centering
\caption{Infection intensity (infections/day) by longevity band}
\label{tab:longevity_intensity_summary}
\begin{tabular}{lcccc}
\toprule
\textbf{Longevity Band} & \multicolumn{2}{c}{\textbf{Software}} & \multicolumn{2}{c}{\textbf{Gaming}} \\
\cmidrule(lr){2-3} \cmidrule(lr){4-5}
 & n & Mean (SD) & n & Mean (SD) \\
\midrule
0–7 days & 33 & 3.48 (2.00) & 5 & 2.87 (2.44) \\
8–14 days & 20 & 1.61 (1.28) & 6 & 0.72 (0.38) \\
15–28 days & 30 & 0.66 (0.56) & 16 & 0.80 (0.52) \\
29+ days & 45 & 1.15 (1.20) & 22 & 2.28 (2.51) \\
\midrule
\textbf{Total} & 128 & --- & 49 & --- \\ 
\bottomrule
\end{tabular}
\end{table}

\begin{table}[ht]
\centering
\caption{Longevity--intensity relationship statistics}
\label{tab:longevity_intensity_models}
\small
\setlength{\tabcolsep}{4pt}
\begin{tabular}{lccccc}
\toprule
\textbf{Type} & \textbf{$r$} & \textbf{$\rho$} & \textbf{Model} & \textbf{$R^2$} & \textbf{$\Delta$AIC\textsuperscript{a}} \\
\midrule
Software (128) & $-0.467$*** & $-0.489$*** & Power & 0.558*** & 6.4 \\
Gaming (49)    & 0.204       & 0.277       & Quad. & 0.337*** & 0.9 \\
\quad $>$7 days\textsuperscript{b} (44) & 0.471** & 0.510*** & Power & 0.261*** & 0.3 \\
\bottomrule
\end{tabular}

\vspace{2pt}
\parbox{\columnwidth}{\footnotesize $r$: Pearson; $\rho$: Spearman; $n$ in parentheses. *** $p<0.001$, ** $p<0.01$, * $p<0.05$.  \textsuperscript{a}AIC difference to the second-best model; values $<2$ indicate models are not meaningfully distinguishable. \textsuperscript{b}Sensitivity analysis excluding the five gaming campaigns lasting $\leq$7 days.}
\end{table}

\begin{figure*}[t] 
\centering 
\includegraphics[width=\textwidth]{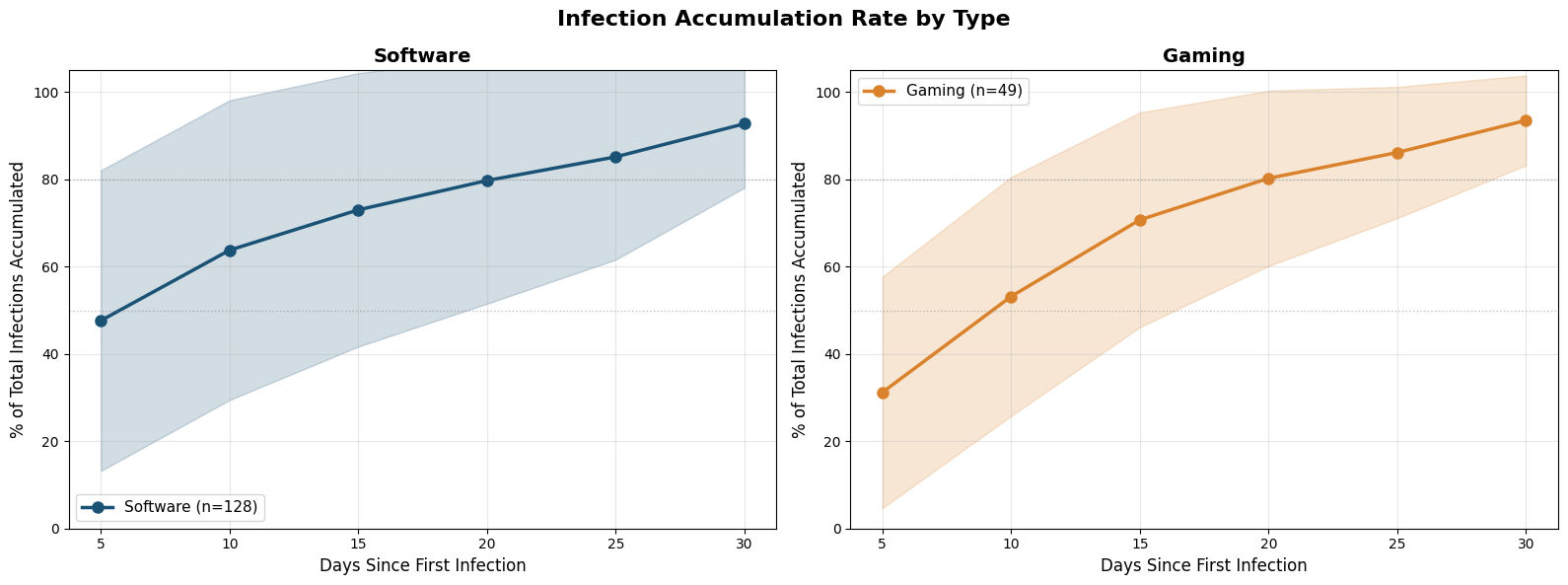}
\caption{Percentage of total infections after x days since first infection.}
\label{fig:accumulation_spe}
\end{figure*}

\paragraph{Infection Accumulation}
\label{sec:accumulation}
The intensity patterns above are mirrored in cumulative infection profiles. Figure~\ref{fig:accumulation_spe} illustrates the temporal accumulation of infections across software and gaming campaigns, while Table~\ref{tab:infection_metrics} summarizes key accumulation milestones. Infection accumulation curves are normalized to each campaign’s observed total population, allowing comparison of temporal shape rather than absolute volume.

Software campaigns accumulate infections rapidly: on average campaigns reach 47.6\% of their total observed infections within five days, 63.8\% by day ten. This behavior is consistent with a \emph{release window} effect, where the publication of a newly cracked or updated version generates a short-lived surge in demand and rapid initial spread. 

Gaming campaigns accumulate more gradually. Only 31.2\% of observed infections occur within the first five days, 53.2\% by day ten. Unlike cracked software, gaming tools such as executors or mod menus do not follow a predictable release cycle. Their adoption often depends on community attention, which can shift at any time. As a result, infections grow more steadily rather than peaking immediately after release, but can surge unpredictably when a tool/mod gains traction in a gaming community. In these cases, infections may surge during a period of heightened interest, often driven by social platforms or forums, before stabilizing once the “hype” subsides.

The difference is largest at the 80\% threshold: 55.1\% of software campaigns reach this milestone within ten days, compared to just 16.3\% of gaming campaigns (Table~\ref{tab:infection_metrics}). Beyond 15–20 days, accumulation plateaus across both ecosystems, indicating that most observed infections occur within the two-week mark.

\paragraph{Operational Implications}

These findings converge on a clear defensive strategy. For software campaigns, early intervention is critical: the first 10 days capture over 60\% of observed infections, and intensity drops rapidly after. For gaming campaigns, sustained monitoring is required: intensity increases with longevity, so long-lived campaigns tend to become more, not less, impactful.

\begin{table}[h!]
\centering
\caption{Infection accumulation metrics for software and gaming campaigns}
\label{tab:infection_metrics}
\begin{tabular}{lcc}
\hline
 & \textbf{Software} & \textbf{Gaming} \\
\hline
First 5 days & 47.6\% & 31.2\% \\
First 10 days & 63.8\% & 53.2\% \\
Campaigns reaching 50\% by day 10 & 60.9\% & 46.1\% \\
Campaigns reaching 80\% by day 10 & 55.1\% & 16.3\% \\
\hline
\end{tabular}
\end{table}

\subsubsection{Post-Discovery Campaign Growth}

To illustrate the operational relevance of early campaign identification, we examine how selected campaigns evolved after their first detection by our framework. These examples are not used to estimate population-level preventable infections, but rather to demonstrate the magnitude and timing of post-discovery growth observed in practice.

A representative example from the software ecosystem is the KMS \texttt{kmsauto\_1.9.9} campaign, the largest KMS campaign in our dataset, with 183 detected infections (187 in the ground truth) over a 36-day period. The framework detected this campaign during its initial observation window (2025-06-09 to 2025-06-30), at which point only 13 infections had been observed. At the time of identification, the campaign contained eight URLs, of which seven pointed to the same origin, a GitHub repository that was active at the time (taken down since). The remaining 170 infections accumulated after this first detection, indicating that the majority of campaign activity occurred after initial discovery. 

A similar pattern is observed in gaming campaigns, albeit at smaller absolute scale. The ROBLOX \texttt{swift} campaign was detected in its first observation window with 23 infections and remained active for 43 days. First detection yielded 10 URLs of which 6 were leading to a file hosting platform. Following detection, the campaign accumulated an additional 43 detected infections, ultimately reaching 66 detected infections (138 in the ground truth, reflecting the lower completeness of gaming campaigns). While the overall volume is lower than in software campaigns, the relative post-discovery growth is comparable.

Across both examples, campaign discovery occurred early relative to total observed lifetime, yet substantial infection growth followed. On average, software campaigns accumulate 43.50\% of their total observed infections after discovery, compared with 29.89\% for gaming campaigns. Although the absolute number of preventable infections depends on campaign completeness and external response capacity, these cases illustrate that early identification often precedes a significant fraction of overall campaign activity. When early detection surfaces valid URLs and indicators of compromise (IoCs), campaigns can be documented by anti-virus vendors and/or disrupted through rapid takedown, potentially preventing a meaningful portion of subsequent infections.

\section{Discussion}

In this section, we discuss our findings and their implications. We examine a) framework performance and its limitations at different campaign scales, b) the robustness of lure-based clustering to infrastructure fragmentation, c) the advantages of a victim-centric approach over traditional detection, d) intervention windows revealed by campaign temporal dynamics, e) ecosystem-specific remediation strategies, and f) how cataloging successful lures enables a preventive awareness paradigm that complements existing technical defenses.

\paragraph{Framework Performance}
Our dual-framework approach achieves strong detection performance for medium and large-scale campaigns, which account for the majority of infections and harm. Reliable detection at this scale makes the framework a useful tool for tracking the most impactful threats. The software framework identified all ground truth campaigns above ten infections across Adobe and KMS brands, though performance degrades for smaller campaigns. The gaming framework detected all large and medium campaigns for ROBLOX and GTA (although Velocity was only recovered merged with Swift), but only 58.33\% (14/24) of small and emerging campaigns, indicating reduced reliability at smaller scales.


\paragraph{Social Engineering Lures as Campaign Identity Brings Robustness to Infrastructure Fragmentation}
Our analysis confirms that social engineering lures serve as reliable campaign signatures for infostealers. Across both software and gaming ecosystems, campaigns are defined not by technical novelty or payload variants, but by what they masquerade as, namely cracked software or gaming enhancements. 

A critical advantage of lure-based clustering is its ability to maintain campaign identity across disjoint infrastructure. For example, The Pirate Bay alone was reached through 20 different mirror hostnames, and gaming campaigns were spread across 229 distinct standalone domains, most used by a single campaign. Infrastructure-based clustering would fragment these campaigns across dozens of unrelated indicators, whereas the lure (e.g., "Kiddions mod menu" or "KMSAuto 1.9.9") remains stable across all of them. This enables campaign identification even when infrastructure is ephemeral, fragmented, or dispersed across legitimate platforms, scenarios where traditional infrastructure-based detection would fail.

\paragraph{Victim-Centric Approach and Advantages}
Unlike prior campaign detection approaches, our framework derives campaign structure directly from victim-side artifacts. By leveraging infostealer logs, our framework captures why campaigns succeed rather than just how they operate. 

This victim-centric perspective remains robust even as threat actors rotate domains, hosting providers, or payload variants, preserving campaign visibility despite infrastructure churn. In practice, attackers could either infect fewer victims or stop sharing stealer logs to limit the applicability of our framework. But both would directly undermine their operational and economic incentives. Overall, by grounding campaign identity in successful infections and leveraging social engineering lures as campaign signatures, our framework complements traditional detection methods. It provides durable, actionable insight into attacker strategy that is otherwise obscured by infrastructure-level analysis alone.

\paragraph{Intervention Windows and Temporal Dynamics}
Our analysis of campaign infection dynamics reveals that early detection creates prevention opportunities. Post-discovery growth patterns show that campaigns typically accumulate 30-44\% of their total infections post first discovery window, meaning that timely intervention can prevent a good portion of harm even when detection is not immediate.

For software campaigns, the first ten days following release are critical: over 60\% of total infections occur during this initial period. This concentrated early activity suggests that rapid detection and takedown of newly released cracks should be prioritized. 

Gaming campaigns present a different temporal profile. Rather than decaying, their intensity increases with longevity among campaigns lasting more than a week: long-running campaigns reach 2.28 infections/day, nearly twice the rate of long-running software campaigns. A gaming campaign that is still active after several weeks is therefore not fading but often gaining momentum. Intervention remains valuable throughout a campaign's lifetime, and more than an early one-time response, gaming campaigns require continuous monitoring.

\paragraph{Infrastructure Reuse and Remediation}

The two ecosystems offer different points of intervention, and the distinction between shared platforms and standalone sites is key to understanding them. Standalone websites are common infrastructure in both ecosystems, but they are largely single-use: taking them down affects one campaign at a time, and threat actors can easily replace them. Shared platforms, by contrast, recur across many campaigns and are where remediation scales. 

For gaming campaigns, these chokepoints are mainstream services such as GitHub, MediaFire, and YouTube, each present in about third or more of campaigns. Because these platforms already operate abuse processes, platform-level detection offers the most efficient lever against gaming campaigns, while takedowns of lookalike standalone sites remain a slower, campaign-by-campaign effort. 

Software campaigns present a different remediation landscape. While fewer campaigns share the same infrastructure, those that do often rely on centralized platforms specializing in cracked software distribution. Tracing a single software campaign could therefore expose infrastructure used by many other malicious distributions. This concentration does not translate into an easy takedown target: piracy platforms are resilient by design, as illustrated by the 20 mirrors through which The Pirate Bay is reached. Their value lies instead in monitoring: observing a small number of torrent indexes provides early visibility into a large share of software campaigns.

\paragraph{Actionability and Preventive Awareness}

Although actionability varies between software and gaming campaigns, early visibility is valuable in all cases. For actionable campaigns, early detection enables direct intervention, such as disrupting infrastructure or sharing IoCs with antivirus vendors to prevent further infections. 

Critically, cataloging successful infection lures enables a prevention paradigm that existing technical defenses cannot provide. While antivirus, blacklists, and network filters can be socially engineered into irrelevance, awareness of specific lures that have proven effective in real-world infections empowers users to recognize deception patterns. For example, knowing that "Adobe Photoshop v25.3.1.122" or "ROBLOX Lua executor" are active lures allows users to make informed decisions before infection occurs. This awareness-based prevention complements technical controls and addresses the human decision-making dimension that traditional campaign detection overlooks.

\section{Limitations and Methodological Considerations}
\label{sec:limitations}

\paragraph{Parameter Sensitivity and the Completeness-Coverage Trade-off}
Experimentation with window size parameters revealed an important trade-off between campaign completeness and campaign coverage. Adjusting the observation window to 14 days with a 1-day step size improved campaign completeness but reduced overall campaign coverage, primarily by failing to identify additional medium and small campaigns. In contrast, larger window sizes improved campaign discovery at the expense of completeness.

This completeness-coverage trade-off has practical implications for operational deployment. Teams prioritizing comprehensive reconstruction of known campaigns may benefit from tighter window parameters. Conversely, teams focused on broad situational awareness and early warning may prefer larger windows that maximize campaign coverage at the cost of some attribution precision.

\paragraph{Sample and Temporal Constraints}
The dataset contains infections observed between June 7th and August 12th, 2025, but 99\% of records are dated before 25 July. Campaigns were therefore identified from infections between June 7th and July 25th 2025, a 47-day window. All longevity measures are bounded by this window. Campaigns already active before June 7th, or still active after July 25th, are truncated, so their lifespans are underestimated. The maximum observed lifespan (44 days) is close to the window length, which suggests that some campaigns lasted longer than we could observe. Reported lifespans should therefore be read as lower bounds, and the analysis is based on 177 campaigns, which limits how far the results can be generalized to the wider infostealer ecosystem.

\paragraph{Completeness Bias}
As established in Section~\ref{sec: completeness}, mean population completeness is 94.56\% for software and 71.13\% for gaming campaigns. Gaming campaign metrics, including infection counts, intensity rates, and temporal patterns, represent conservative lower-bound estimates rather than exhaustive reconstructions. True gaming campaign activity is likely approximately 1.4 times higher than observed values.

Importantly, this underestimation does not undermine our conclusions; if anything, it strengthens them. The finding that gaming campaigns require sustained monitoring holds even more strongly if true intensity exceeds our measurements. Nonetheless, readers should interpret absolute gaming metrics with this bias in mind.

\paragraph{Data Source Limitation}
Low software URL density may reflect data artifacts rather than true infrastructure characteristics. Our analysis relies on screenshot data captured during infection events. For software campaigns, screenshots are typically taken during the launch of fake installers, showing only the installer interface rather than the webpage from which the download was obtained. URL extraction will drastically improve by including all artifacts (i.e. browsing history and software installed) from the log into the infection theme analysis pipeline. Future iterations of the framework integrating these artifacts could enable broader IoC collection and more effective remediation.

\paragraph{Pipeline-Specific Limitations}

Product misnaming in the software framework is the most harmful since it cascades through all downstream analysis. One possible mitigation is to use stricter initial clustering to create smaller, more precise groups, accepting a higher number of clusters in exchange for better label accuracy. These clusters can then be merged at the product grouping stage. Additionally, applying semantic isolation scoring to software campaign discovery could improve author-based detection, as author and repackager pseudonyms, like gaming campaign identifiers, tend to be semantically isolated.

The gaming framework operates on the assumption that campaigns use identifiable naming patterns, typically executor or cheat tool names that serve as natural campaign identifiers. This assumption holds for games with established cheating culture (ROBLOX, GTA, Fortnite) where executor branding is common. However, games where campaigns are named after in-game features rather than tools would require substantial generic word list modification. Games lacking an executor culture entirely would reduce the framework to a more manual process, undermining its automation benefits.

For both pipelines, the generic word list curation process represents a practical limitation. Effective campaign identification requires an exhaustive list of brand-specific generic terms and creating such lists requires significant domain knowledge, is very time-consuming and must be repeated for each brand. Finally, hyperparameters also differ across brands, requiring brand-specific tuning. This dependency on manual curation limits the framework's scalability and introduces potential for inconsistency across brands.

\section{Conclusion}
This research introduces a dual-framework capable of identifying and tracking infostealer campaigns and related IoCs at scale in near real-time. Recognizing that software and gaming campaigns exhibit fundamentally different naming patterns, we developed pipelines tailored to each ecosystem. Despite limitations in detecting small and emerging gaming campaigns, the framework reliably identifies and tracks medium and large-scale campaigns, which account for the majority of infections.

Our analysis of identified campaigns reveals distinct behavioral and infrastructure signatures across ecosystems, with direct remediation implications. Software campaigns demand rapid response: intervention within the first ten days of a campaign captures the majority of preventable infections, and monitoring a small number of torrent indexes covers a large share of software campaigns. Gaming campaigns reward sustained monitoring, as their intensity grows with longevity, and remediation requires both cross-platform abuse reporting and takedowns of campaign-specific lookalike domains.

More broadly, this work argues for a shift in how infostealer campaigns are identified and studied. By grounding campaign identity in victim-side artifacts rather than infrastructure, this framework captures why campaigns succeed, not just how they operate. Social engineering lures function as stable campaign signatures that provide robustness to infrastructure fragmentation, conditions where traditional infrastructure-based detection fails.

Where prior work analyzes post-infection artifacts in isolation, our framework structures this data into coherent campaign-level intelligence. This enables systematic campaign-specific IoC collection, distribution infrastructure analysis, and campaign effectiveness measurement, providing a practical foundation for scalable detection, remediation, and awareness-driven mitigation of infostealer threats.

\bibliographystyle{plain}
\bibliography{lit_review}

\section*{Appendix}
\label{app:pseudocode}

This appendix provides the complete algorithmic specification for our 
Hybrid Rolling HDBSCAN implementation described in 
Section~\ref{sec:hybrid_hdbscan}, and Table~\ref{tab:parameters} lists the HDBSCAN parameters used per ecosystem.

\begin{figure*}[p]
\centering
\begin{adjustbox}{max width=\textwidth, max totalheight=0.9\textheight}
\begin{tikzpicture}[node distance=1cm and 2cm, auto]

\node (start) [startstop] {START};

\node (init) [process, below=0.5cm of start] {Initialize:\\$\mathcal{C}_{active} = \emptyset$, $id = 0$, $w = 0$};

\node (extract) [process, below=0.5cm of init] {Extract Window $\mathcal{W}_w$\\Get embeddings $\mathbf{E}_w$};

\node (check_window) [decision, below=0.5cm of extract] {$w = 0$?};

\node (window0) [window0, below left=1.2cm and 2.5cm of check_window] {
    \textbf{Window 0: Initial Training}\\[2pt]
    1. Train HDBSCAN on $\mathbf{E}_0$\\
    2. Compute centroids\\
    3. Calculate $\sigma$\\
    4. Mark as NEW
};

\node (windowN) [windowN, below right=1.2cm and 2.5cm of check_window] {
    \textbf{Window $n>0$: Incremental}\\[2pt]
    1. approximate\_predict()\\
    2. Separate assigned/noise\\
    3. Update clusters\\
    4. Recluster noise
};

\node (merge1) [merge, below=3cm of check_window] {};


\node (retrain_decision) [decision, below=1.5cm of merge1] {Retraining Conditions?};

\node (full_retrain) [retrain, below left=1cm and 3cm of retrain_decision] {
    \textbf{Full Retrain}\\[2pt]
    1. Store old\\
    2. Retrain HDBSCAN\\
    3. Hungarian match\\
    4. Preserve IDs\\
    5. Recompute $\sigma$
};

\node (surgical) [retrain, below=1cm of retrain_decision] {
    \textbf{Surgical}\\[2pt]
    1. Find drifted\\
    2. To noise\\
    3. Recluster\\
    4. Preserve stable
};

\node (continue) [retrain, below right=1cm and 3cm of retrain_decision] {
    \textbf{Continue}\\[2pt]
    No significant\\
    drift detected.\\
    Use current\\
    structure.
};

\node (merge2) [merge, below=3cm of retrain_decision] {};

\node (lifecycle) [lifecycle, below=1.5cm of merge2] {
    \textbf{Update Lifecycle}\\[2pt]
    Points? $\rightarrow$ ACTIVE\\
    No points? $\rightarrow$ DORMANT\\
    Revived? $\rightarrow$ REVIVED
};

\node (remove_expired) [lifecycle, below=0.75cm of lifecycle] {
    \textbf{Remove Expired}\\[2pt]
    dormant\_count $> M_{dormant}$\\
    $\rightarrow$ EXPIRED $\rightarrow$ DELETE
};

\node (slide) [process, below=of remove_expired] {Slide Window:\\$w \gets w + 1$};

\node (check_data) [decision, below=0.75cm of slide] {More\\data?};

\node (output) [output, below=0.75cm of check_data] {Output:\\Campaigns};

\node (end) [startstop, below=0.75cm of output] {END};

\draw [arrow] (start) -- (init);
\draw [arrow] (init) -- (extract);
\draw [arrow] (extract) -- (check_window);

\draw [arrow] (check_window) -- node[arrow_label, above left, pos=0.3] {Yes} (window0);
\draw [arrow] (check_window) -- node[arrow_label, above right, pos=0.3] {No} (windowN);

\draw [arrow] (window0) |- (merge1);
\draw [arrow] (windowN) |- (merge1);

\draw [arrow] (merge1) -- (retrain_decision);

\draw [arrow] (retrain_decision) -- node[arrow_label, above left, pos=0.25] {$> P_{\text{drift}}$ or $> P_{\text{dissolve}}$ or large dissolution} (full_retrain);
\draw [arrow] (retrain_decision) -- node[arrow_label, left, pos=0.3] {$< P_{\text{drift}}$} (surgical);
\draw [arrow] (retrain_decision) -- node[arrow_label, above right, pos=0.25] {No} (continue);

\draw [arrow] (full_retrain) |- (merge2);
\draw [arrow] (surgical) -- (merge2);
\draw [arrow] (continue) |- (merge2);

\draw [arrow] (merge2) -- (lifecycle);
\draw [arrow] (lifecycle) -- (remove_expired);
\draw [arrow] (remove_expired) -- (slide);
\draw [arrow] (slide) -- (check_data);

\draw [arrow] (check_data.east) -- node[arrow_label, above] {Yes} ++(6,0) |- (extract.east);
\draw [arrow] (check_data) -- node[arrow_label, left] {No} (output);
\draw [arrow] (output) -- (end);

\begin{scope}[on background layer]
    \node[draw=myblue2, thick, rounded corners, 
          fill=myblue, fill opacity=0.3,
          fit=(window0) (windowN) (merge1), 
          inner sep=8pt, 
          label={[shift={(5.2,0.1)}]:\scriptsize\textbf{WindowProcessing}}] {};
    
    \node[draw=retraincolor2, thick, rounded corners, fill=retraincolor, fill opacity=0.1,
          fit=(retrain_decision) (full_retrain) (surgical) (continue) (merge2), 
          inner sep=8pt, label={[shift={(5.2,0.1)}]:\scriptsize\textbf{Adaptive Retraining}}] {};
    
    \node[draw=lifecyclecolor2, thick, rounded corners, fill=lifecyclecolor,  fill              opacity=0.3, fit=(lifecycle) (remove_expired), 
          inner sep=8pt, inner xsep=2cm, label={[shift={(2.2,0.1)}]:\scriptsize\textbf{Lifecycle Management}}] {};
\end{scope}

\end{tikzpicture}
\end{adjustbox}
\caption{Complete vertical workflow of the Hybrid Rolling HDBSCAN algorithm.}
\label{fig:full_workflow_vertical}
\end{figure*}
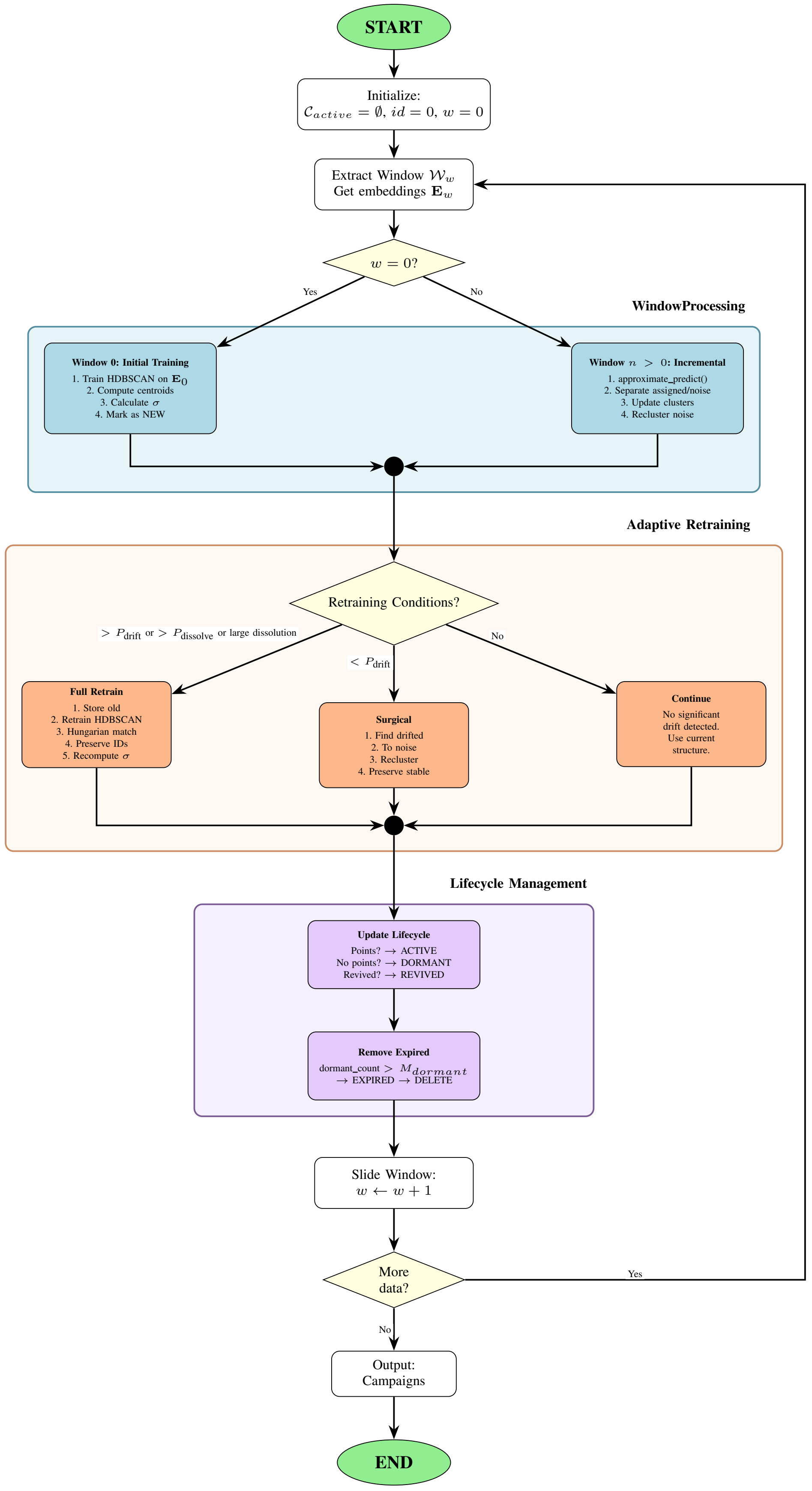

\subsection*{Parameter Guidelines}

\begin{table}[h]
\centering
\caption{HDBSCAN parameters by ecosystem}
\label{tab:parameters}
\begin{tabular}{lcc}
\toprule
\textbf{Parameter} & \textbf{Gaming} & \textbf{Software} \\
\midrule
$\theta_{\text{min\_size}}$ & $[5:10]$ & $[5:10]$ \\
$\theta_{\text{min\_samples}}$ & $[5:10]$ & $[5:10]$ \\
$\tau_{\text{drift}}$ & 2.2 & 2.2 \\
$\tau_{\text{match}}$ & 1.0 & 1.0 \\
$\varepsilon$ & $0.25$ & $[0.1:0.2]$ \\
\bottomrule
\end{tabular}
\end{table}

\end{document}